\documentstyle[eqsecnum,aps,prd,preprint,epsfig]{revtex}
\begin{document}


\def\bq{\begin{equation}}
\def\nq{\end{equation}}
\def\bqr{\begin{eqnarray}}
\def\nqr{\end{eqnarray}}
\let\l=\left
\let\r=\right


\title{Modelling the Dynamics of Global Monopoles}   
\author{Inyong Cho\footnote[1]{Electronic address: cho@cosmos2.phy.tufts.edu}
and  Jemal Guven\footnote[2]{Electronic address: jemal@nuclecu.unam.mx}
\footnote[3] {Permanent Address:
 Instituto de Ciencias Nucleares,
 Universidad Nacional Aut\'onoma de M\'exico.
 Apdo. Postal 70-543, 04510 M\'exico, D.F., MEXICO}}

\address{Institute of Cosmology,
        Department of Physics and Astronomy,\\
        Tufts University,
        Medford, Massachusetts 02155, USA}
\date{\today}
\maketitle

\begin{abstract}
A thin wall approximation is exploited to describe
a global monopole coupled to gravity. The core is modelled by
de Sitter space; its boundary by a thin wall with a 
constant energy density; its exterior by the asymptotic 
Schwarzschild solution with negative gravitational mass $M$
and solid angle deficit, $\Delta\Omega/4\pi
= 8\pi G\eta^2$, where $\eta$ is
the symmetry breaking scale. The deficit angle equals $4\pi$
when $\eta=1/\sqrt{8\pi G} \equiv M_p$.
We find that: (1) if $\eta <M_p$, there exists a unique 
globally static non-singular solution 
with a well defined mass, $M_0<0$. $M_0$ 
provides a lower bound on $M$. If $M_0<M<0$, the
solution oscillates. There are no inflating solutions in this 
symmetry breaking regime.
(2) if $\eta \ge M_p$, non-singular
solutions with an inflating core and  
an asymptotically cosmological exterior will exist 
for all $M<0$. 
(3) if $\eta$ is not too large, there exists a finite 
range of values of $M$ where a 
non-inflating monopole will also exist. These
solutions appear to be metastable towards inflation.
If $M$ is positive all solutions are singular.
We provide a detailed description of the 
configuration space of the model for each
point in the space of parameters, $(\eta, M)$
and trace the wall trajectories on both the 
interior and the exterior spacetimes.
Our results support the proposal that
topological defects can undergo inflation.

\end{abstract}

\section{Introduction}

Phase transitions occuring in the early universe can give rise to topological
defects of various kinds; these may be domain walls, strings and monopoles
as well as more exotic objects\cite{VS}. 
Recently Vilenkin \cite{AV} and Linde \cite{AL} proposed that topological 
defects could inflate when the symmetry breaking scale,
$\eta \gtrsim {\cal O}(m_p)$, thereby providing natural seeds 
for an inflating universe. This was recently confirmed
numerically by Sakai {\it et al.} \cite{NS} who showed that
domain walls and global monopoles will inflate if $\eta \gtrsim 0.33m_p$. 
However, until now, there has 
been no analytical confirmation of their results.

The simplest defects  
are global monopoles which are localized in all three 
spatial directions. 
Barriola and Vilenkin \cite{BV} obtained the simplest static 
global monopole solution coupled to gravity.
Asymptotically, the  spacetime is described 
by the static line element,

\bq
        ds^2 = -A_M dT_M^2 + A_M^{-1} dR^2 + R^2 d\Omega^2\,, 
\label{eq:EXT}
\nq
where

\bq
        A_M = 1-8 \pi G \eta^2 - {2GM \over R}\,.
\label{eq:AM}
\nq
The gravitational mass is given by the parameter, $M$.
There are several features of this asymptotic spacetime which are 
unusual:

At infinity, this spacetime is not flat.
There is a solid angle deficit, $\Delta/4\pi \simeq 8\pi G\eta^2$, 
determined completely by the symmetry breaking scale, $\eta$.
The occurence of a deficit angle is a consequence of the non-trivial topology 
of the field configuration outside  the monopole:
non-vanishing gradients  along non-radial directions
give an energy density outside which falls off slowly, $\sim \eta^2/R^2$. 
Eq. (\ref{eq:EXT}) is then the most general spherically symmetric
solution of the Einstein equations consistent
with such a  source. This is analogous to a global string\cite{Greg}.

As Harari and Lousto pointed out, the mass parameter in this 
static metric is always negative \cite{HL}. This is not, however,  
a violation of the positive mass theorem ---  the spacetime is 
not asymptotically flat so that the gravitational mass
does not coincide with the ADM mass at infinity\cite{Sud}. 

A consequence of the slow falloff is that 
the total energy of the monopole diverges linearly. 
To regularize this energy we need to introduce a cutoff at some
large radius, $R^*$. This cutoff will be provided 
by the correlation length of the scalar field, $\xi$.
In cosmology, an upper bound on $\xi$ is provided by the 
horizon size. 

When the solid angle deficit exceeds $4\pi$, 
($8\pi G\eta^2 >1$),
the roles of $R$ and $T_M$ get interchanged. 
The exterior solution which corresponds to 
Barriola and Vilenkin's ansatz is no longer static.
This is precisely the regime where topological 
inflation is predicted to occur and is the regime we will be
particularly interested in.

In this paper, we present a model of a 
global monopole which is 
tractable analytically and, we believe, 
includes all of its essential features. The core of the global monopole
is approximated by a spherically symmetric 
region of false vacuum, with energy density $\rho$, and 
radius $r$. Outside this core is described by 
a spherically symmetric region with energy density
$\eta^2/R^2$ and thus described by the 
the asymptotic static metric (\ref{eq:EXT}).

If the approximation stops here, a static equilibrium exists 
between these two spacetime geometries
only when 

\bq
r= {\eta\over \sqrt{\rho}}
\,,\label{eq:HL1}
\nq  
and the mass assumes the 
negative value, 
\bq
M = - {8\pi\over3} {\eta^3\over \sqrt\rho}
\,.\label{eq:HL2}
\nq
For a given theory this solution is unique. However, the 
model suffers from the shortcoming that it 
only describes a static equilibrium  and  predicts
such an equilibrium for all values of $\eta$;
it does not possess 
the scope to describe non-static configurations.
This is essentially because the energy density of 
false vacuum is constant. 
In general, when false vacuum is converted to true 
vacuum (with constant solid 
angle deficit), the energy released  
is transferred  to the core boundary \cite{Cole}. 
This boundary plays an essential 
dynamical role. The necessary refinement of the model is to
introduce 
a surface layer with energy density, $\sigma$, on the core boundary. 
On dimensional grounds, we expect $\rho\sim \eta^4$ and
$\sigma\sim \eta^3$. 
The Einstein equations now determine the motion of $r$.

If $\eta< M_p$\footnote[1] {For convenience, we introduce $M_p \equiv m_p/\sqrt{8\pi}$.},
we find stable oscillating
solutions for each negative value of $M$ above some  
threshold. 
These solutions are the analogues of Harari and Lousto's static approximation.

If $\eta\ge M_p$, non-singular
solutions with inflating cores and asymptotically cosmological 
exteriors exist for all $M<0$.  
There exists some critical value of $\eta$, $\eta_c > M_p$ above which
all monopoles inflate.

For each $\eta$ in the interval, $M_p <\eta < \eta_c$
a stable oscillating monopole will co-exist with an
inflating one  of the same mass in some  
strictly negative band of values of $M$. 

These solutions have the virtue that they are non-singular 
everywhere.

If $M> 0$, all solutions 
collapse to form a black hole.
There also exist collapsing monopole solutions with $M<0$
which terminate in a naked singularity. 
However, they do not possess a foliation as an isolated object in 
an asympotically cosmological spacetime. 
We therefore dismiss them as unphysical.

We examine the global 
geometry of the corresponding monopole spacetimes when $M<0$.
When $\eta >M_p$, the exterior spacetime possesses a 
cosmological horizon. We construct explicitly the 
Kruskal-Szekeres coordinate system which
is non-singular on this horizon. 
The maximal extension of the 
exterior geometry is then presented.
To provide a physical description of the exterior spacetime 
we identify explicitly a foliation of this geometry
which corresponds to an isolated object in an
asymptotically cosmological spacetime.

The paper is organized as follows:
In Sec. II, we write down the Einstein equations  
for our model.
In Sec. III, the simpler problem 
of wall motion is Minkowski space is discussed.
In Sec. IV, we examine the motion of the wall coupled to gravity
and describe all possible trajectories.
In Sec. V, we trace the wall trajectory in spacetime.
In Appendix A, the wall equation of motion is analysed in detail in a 
simple tractable special case. 
In Appendix B, the routing of the wall trajectories on
Kruskal-Szekeres and Gibbons-Hawking diagrams is determined.

\section{Einstein's Equations at the Wall}

The simplest model that admits global monopoles 
is described by the Lagrangian 

\[
        {\cal L} = -{1 \over 2}\partial_\mu\phi^a\partial^\mu\phi^a 
        -{1 \over 4} \lambda (\phi^a\phi^a - \eta^2)^2\,,
\]
where $\phi^a$ is a triplet of scalar fields.
The ansatz which describes a static monopole with unit 
topological charge is
(as in Minkowski space) $\phi^a = \phi (R)\, \hat x^a$,
where $\hat x^a$ is a radial unit vector. 
The corresponding spacetime is spherically symmetric.  

At the center of the monopole, the field is in the false vacuum,
$\phi=0$, with energy density
$\rho={1 \over 4} \lambda\eta^4$ and pressure $P=-\rho$.  
We will approximate the core of radius $r$ by a region of 
false vacuum. The interior spacetime is then 
de Sitter space, which we can describe by the static line element,

\bq
        ds^2 = -A_D dT_D^2 + A_D^{-1} dR^2 + R^2 d\Omega^2\,, 
\label{eq:DS}
\nq
where 
        
\bq
        A_D = 1 - H^2R^2 \,, \qquad
        H^2 = {8 \pi G \over 3} \rho\,.
\label{eq:H}
\nq
The static chart will describe the interior if 
$HR <1$ everywhere.  

Asymptotically, $\phi \approx \eta $. The stress 
tensor assumes the perfect fluid form with $P=-\rho_{\rm ext} (R)$
where $\rho_{\rm ext}(R)\approx \eta^2/ R^2$. The 
corresponding spacetime is 
described by the line element (\ref{eq:EXT}).
Harari and Lousto's numerical calculations 
show that the asymptotic form (\ref{eq:EXT}) is 
approached rapidly outside the core.
We will approximate the solution everywhere 
outside the core by this asymptotic form.

The exterior geometry will be static only if
$\eta <M_p$. When this limit is breached,
the region $R > -2M/(\eta^2/M_p^2 -1)$ will assume a dynamical character. 
Strictly speaking, this is 
inconsistent with the static ansatz we exploit to generate the 
asymptotic form of the metric. To resolve this 
inconsistency we must re-interprete the original ansatz 
as a description of a dynamical object. What has happened is that 
the timelike Killing vector which characterizes a static geometry
has become spacelike. The 
radial vector, however, remains normal to the Killing vector.
The cosmological nature of the exterior geometry 
will be described more fully in Sec. V.

The continuity of the lapse and its first derivative across the 
core boundary
determines the equilibrium values  (\ref{eq:HL1}) and (\ref{eq:HL2}).
We note that the maximum core size 
covered by a static chart is given by 
$HR = 1$. This limit obtains when $\eta=\sqrt{3} M_p$.
We conclude that a static interior is 
possible well into the regime $\eta >M_p$.

Now let us include a surface energy density on the core boundary.
In the thin wall approximation,  
we can exploit Gaussian normal 
coordinates adapted to the wall to express  the stress tensor 
there in the form 

\bq
        T^a{}_b  = \sigma\delta(n)\delta^a{}_b\,.
\label{eq:Tab}
\nq
$\sigma$ is the constant surface energy density of the boundary. 
The parameter $n$
appearing in Eq. (\ref{eq:Tab}) is the proper distance normal to the 
worldsheet of the wall.  The metric induced on the wall is given by

\[
        ds^2 = -d\tau^2 + r(\tau)^2 d\Omega^2, 
\]
where $\tau$ is the proper time registered 
by an observer at fixed $\theta$ and $\varphi$ who moves with the wall.
The problem reduces to the determination of the 
trajectory, $r=r(\tau)$.

The origin of the
surface energy density are the field gradients 
interpolating between the false vacuum interior
and the exterior. We approximate it

\bqr
        \sigma  &\sim & r_w \times
                        {\rm energy\; density\; difference} \nonumber\\
                &\sim & r_w\,(\rho - \eta^2/r_w^2) \nonumber
\nqr
where $r_w$ is the size of the monopole.
The numerical results of Ref.\cite{HL,NS} show that $r_w$
is proportional to $1/\eta$.  The above relation becomes

\bq
        \sigma \simeq s \eta^3\,,
\label{eq:sigma}
\nq
where $s$ is a dimensionless constant.

The surface energy distribution (\ref{eq:Tab}) introduces a discontinuity  
in the spacetime metric at the wall.
The Einstein equations at the core boundary
reduce to the form \cite{IS,BGG,FGG}
\bq
        K^a{}_b({\rm out})- K^a{}_b({\rm in}) = 
        -4\pi\sigma G\delta^a{}_b\,,
\label{eq:Gauss} 
\nq
where $K_{ab}({\rm in})$ and $K_{ab}({\rm out})$ are
respectively the extrinsic curvature of the wall
embedded in de Sitter space and in the exterior 
spacetime described by (\ref{eq:EXT}). 
Using the techniques developed in \cite{BGG} or \cite{FGG}, 
we find that

\[
        K_{\theta\theta}({\rm in}) = r\beta_D,\qquad
        K_{\theta\theta}({\rm out}) = r\beta_M.
\]
Here, $\beta_D$ and $\beta_M$ are given by
\[
        \beta_D = -A_D \dot{t}_D,\qquad
        \beta_M = A_M \dot{t}_M\,,
\]
where 
$t_D(\tau)$ and $t_M(\tau)$ are  the de Sitter and exterior `static'
time variables, $T_D$ and $T_M$ evaluated on the wall. The overdots refer to 
derivatives with respect to proper time. 
We recall 
that \cite{BGG}

\[
        \beta_D^2  = A_D + \dot{r}^2\,,\qquad 
        \beta_M^2 = A_M + \dot{r}^2\,.
\]
Now the $(\theta\theta)$ component of Eq. (\ref{eq:Gauss}) reads       
\bq
        \beta_D - \beta_M = 4 \pi G \sigma r\,.
\label{eq:Theta}
\nq
This equation can be cast in the form,

\bq
        \dot{r}^2 + U(r) = -1, 
\label{eq:motion}
\nq
where 

\bq
          U(r) = - \left({F_-\over r^2}\right)^2 - H^2 r^2\,,
\label{eq:U-}
\nq
or, alternatively,

\bq
          U(r) = - \left({F_+\over r^2}\right)^2 - {2 GM\over r} - 
8\pi G\eta^2\,,
\label{eq:U+}
\nq
with
\bq
          F_\pm(r) = 
                 {M\over 4\pi\sigma} 
                       -  {\rho_\pm\over 3\sigma} r^3 + 
                             {\eta^2 \over \sigma} r\,, \qquad
          \rho_\pm = \rho \pm  6\pi G \sigma^2\,.
\label{eq:F}
\nq
The linear term in $F_\pm$ 
encodes completely the topology of the scalar field.

\section{Wall Motion in Minkowski Space}

In the limit $G\to 0$, the Einstein equations 
should reproduce the description of a 
global monopole in Minkowski space. 
In this section, we examine  a model of a monopole in Minkowski space. 
This is a useful preliminary step before attempting to examine 
Eq. (\ref{eq:motion}) in all its glory.

Let the core radius be $r$. The energy density in the core is 
a constant, $\rho$, and that in the exterior $\sim \eta^2/R^2$.
A surface layer with energy density, $\sigma$, is 
located on the core boundary.
The total energy of a static configuration is then given by

\bq
        E= {4\pi \over 3} \rho r^3 + 
4\pi \sigma r^2 + 4\pi\eta^2 (R^*-r)\,,
\label{eq:E0}         
\nq
where $R^*$ is a cutoff. 
It is clear that an equilibrium exists. 
We minimize $E$ with respect to $r$ to obtain the stable equilibrium 
core size,

\bq
        r_0= -{\sigma\over \rho} +
              \sqrt{\left({\sigma\over\rho}\right)^2 + {\eta^2\over \rho}}
\,.
\label{eq:r0}
\nq
The (subtracted) energy,
$E^* \equiv E-4\pi\eta^2 R^*$ is then given by 

\bq
        E_0^* = M_0= - {4\pi\over3} {\sigma^2 r_0\over \rho} 
        \left( 1+ 2 {\rho\eta^2\over\sigma^2}
        - \sqrt{1 + {\rho\eta^2\over\sigma^2}}\right)\,.
\label{eq:E1}
\nq 
We note that $M_0$ is manifestly negative.
In the limit  $\sigma\to 0$, we have 
$r_0= {\eta\over \sqrt{\rho}}$ and 
$E_0^* = - {8\pi\over3} {\eta^3\over \sqrt\rho}$.
If we identify $E^*$ with $M$, we reproduce the values obtained by
Harari and Lousto which are independent of $G$.

More generally, let us examine the motion of the core boundary.
The classical action describing this motion  
is given by

\bq
        S  =\int dt \, \left[ {4\pi \over 3} \rho r^3 
        + 4\pi\sigma r^2 \sqrt{1 - \left({dr\over dt}\right)^2}  
                          + 4\pi\eta^2 (R^*-r)\right]\,.
\label{eq:L}
\nq
The canonical conserved energy is given by
a Legendre transformation of the Lagrangian appearing in Eq. (\ref{eq:L}),

\bq
        E^* = {4\pi \over 3} \rho r^3 + 
               4\pi\sigma r^2 \sqrt{1 + \dot r^2}  
             - 4\pi\eta^2 r\,.
\label{eq:E2}
\nq
The only change with respect to Eq. (\ref{eq:E1}) is that the boundary 
term picks up kinetic energy, $4\pi \sigma r^2 \to 
4\pi\sigma r^2 \sqrt{1 + \dot r^2}$.
We can recast Eq. (\ref{eq:E2}) in the form

\bq
        \sqrt{1 + \dot r^2} = {F\over r^2} \,,
\label{eq:sq}
\nq
where $F$ is given by 

\[
        F= {E^*\over 4\pi\sigma} 
                    -  {\rho\over 3\sigma}  r^3  
                            +{\eta^2 \over \sigma} r\,. 
\]
Eq. (\ref{eq:sq}) implies that
$F$ must be positive on any physical trajectory.
This condition places a constraint on the extent of the  
radial domain of the wall.

When $\dot r^2 <<1$, the motion is well described by the 
potential appearing on the RHS of Eq. (\ref{eq:E0}).
Clearly, the equilibrium configuration has $r=r_0$ given by
Eq. (\ref{eq:r0}) with energy, $M_0$, given by Eq. (\ref{eq:E1}).
In general, we can cast Eq. (\ref{eq:sq}) in the form

\[
         \dot r^2 + U(r) = -1\,,
\]
where

\bq
          U(r) = - \left({F\over r^2}\right)^2\,.
\label{eq:U}
\nq

This is exactly the limit $G\to 0$ of Eq. (\ref{eq:motion}) when 
$E^*$ is identified with $M$. We will examine the 
solutions admitted by this system with our eye
on the analogy with the general relativistic problem.

Let us first examine $M<0$.

We first determine where $F$ positive. 
We note that for each negative value of $M$ above some
lower threshold, $M_c$, the cubic function 
$F$ has two positive roots, $r_1$ and $r_2$ say, and  
$F$ is positive only in the domain $[r_1,r_2]$ --- physical motion 
is necessarily bounded. To determine $M_c$, we observe 
that when $M$ falls to 
$M_c$, the two roots coalesce with the local maximum. Thus 
$F=0$ and $F'=0$ simultaneously. We find that this occurs at $r=r_c=
{\eta\over \sqrt{\rho}}$ and  

\bq
         M_c = - {8\pi\over3} r_c \eta^2   \,.
\label{eq:Mc}
\nq
Below $M_c$,  $F$ is negative everywhere. 
Therefore $M =M_c$ places a lower bound on the mass spectrum.

It is now simple to construct the potential, $U$, in the regime $M_c<M<0$.
We note that $r^5 U'= - 2F( r F' - 2 F)$.  
$U$ possesses two maxima  coinciding with the roots of 
$F$ at $r_1$ and $r_2$.
It possesses a single minimum
given by the single positive root of the cubic, $rF'-2 F$. 

On a physical trajectory, $U\le -1$.
The static solution
with $U(r)=-1$ at its minimum determines the 
sharp lower bound, $M_0$ given by Eq. (\ref{eq:E1}), on the mass. 
We note that $r_0 < r_c$. It is then simple to see that $M_c < M_0$
for all $\eta$.
The existence of the threshold, $M_c$, 
places no constraint on the spectrum of oscillating solutions.

For each $M_0 < M <0$, $r$ will oscillate 
between two turning points, $r_{\rm Min}$ 
and $r_{\rm Max}$, bounded within the interval, $[r_1,r_2]$.
This solution will be stable.
If initially the core is displaced from equilibrium
it will oscillate about the equilibrium.
In a physical monopole, we would expect the core to relax to 
$r_0$ as the monopole radiates away its excess energy.

\begin{figure}
\begin{center}
\psfig{file=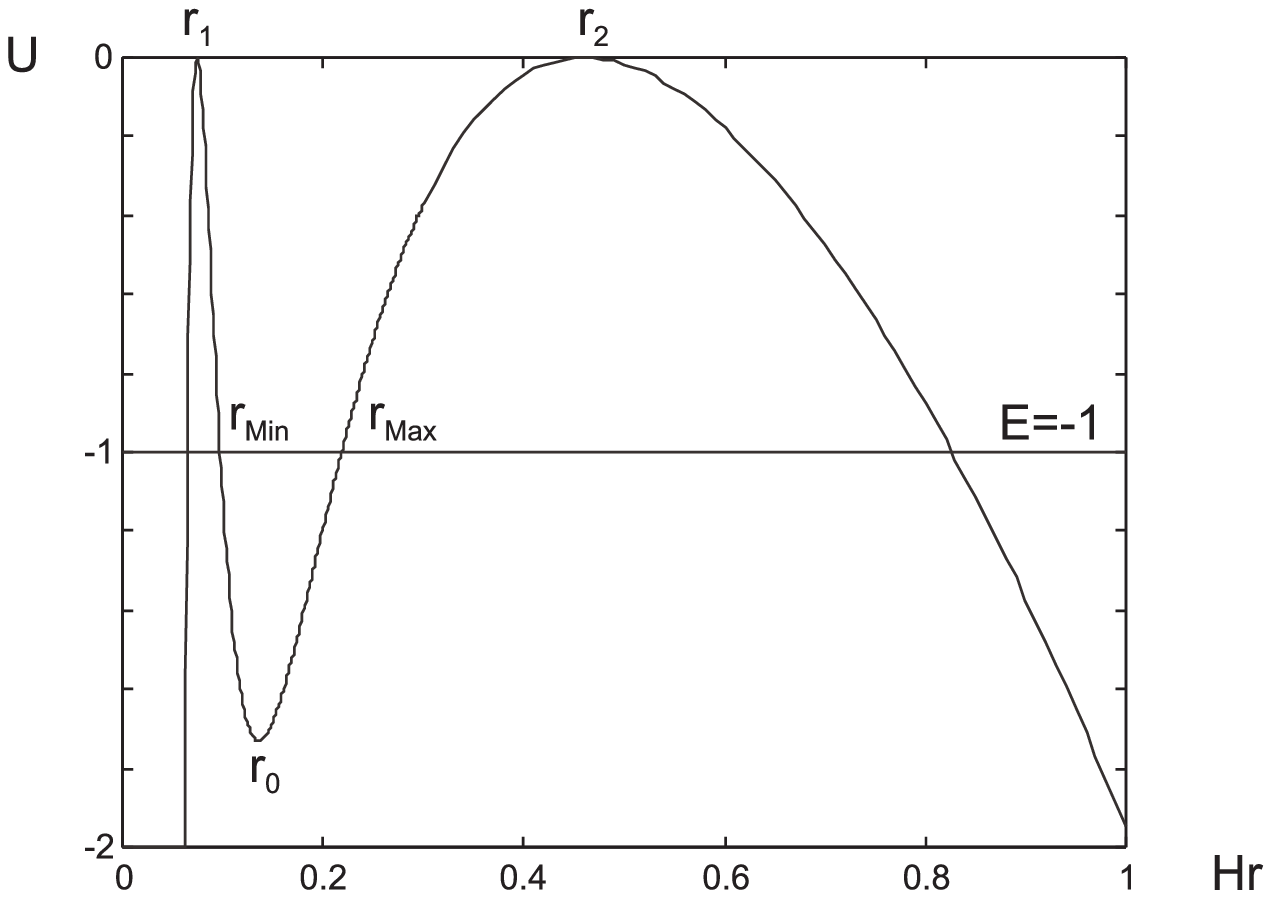}
\end{center}
\caption{
Plot of $U(r)$ vs. $r$  in Minkowski space with
$\eta =0.1m_p$, $M=-2m_p$. }
The wall oscillates between a minimum at $r_{\rm Min}$
and a maximum at $r_{\rm Max}$. 
The physical domain of $U$ is  
contained within the interval, $[r_1,r_2]$, over which
$F$ is positive. 
\label{fig=oscmink}
\end{figure}

The potential $U(r)$ is plotted in Fig. \ref{fig=oscmink} for a
negative value of $M$. 

While both collapsing and expanding solutions appear to exist in the 
region outside the interval, $[r_1,r_2]$, $F<0$ there. Such solutions 
are not consistent with the boundary conditions 
which correspond to a monopole. They correspond instead to 
an unphysical inside-out monopole with an interior with energy density,  
$\eta^2/R^2$ and an exterior false vacuum. It is simple to check that 
$F$ now appears in Eq. (\ref{eq:sq}) with a minus sign.

If $M\ge 0$, $F$ is positive in some finite domain containing the 
origin. This places a bound on the physical domain
of $r$. $U$ increases monotonically  over this domain. 
The core always collapes in a finite proper time, $\tau_0$, with
$r\sim (\tau_0 -\tau)^{1/3}$. 

To summarize, in Minkowski space we saw that the 
spectrum of physically realized values of 
$M$ was bounded from below by $M_0$. For  
each $M_0\le M<0$, there is a stable oscillating solution and
such oscillating solutions exist for all values of $\eta$. 
If $M\ge 0$ there is a bounded solution which collapses.
The only possible motion in Minkowski space 
is bounded. 

\section{Wall Motion}

In this section, we discuss the wall motion in general relativity.
As we will see, when gravity is turned on 
the situation can be quite different.
In addition to determining the 
energy density in the interior and on the core boundary 
$\eta$ also determines the angle deficit in the 
exterior.

We would expect the interior to inflate 
when the core radius exceeds the de Sitter horizon radius.
On dimensional grounds this occurs when $\eta \gtrsim M_p$.
Is this a necessary condition?

If $\eta <M_p$, the deficit $\Delta\Omega <4\pi$.  Thus, if
$M\ge 0$, we would expect to reproduce at least qualitatively, 
the behavior of a false vacuum bubble discussed in \cite{BGG}.  
However, while a false vacuum bubble with $M>0$
always collapses in Minkowski space, when gravity is taken into account
its interior may inflate for any value of the symmetry breaking scale
without destroying the exterior and therefore without any violation of 
flat space intuition in that region.
This is because the inflation occurs behind the event horizon of the 
exterior Schwarzschild geometry.
The inflating interior is connected by a 
wormhole behind this horizon. The 
event horizon signals the eventual formation of a 
black hole. All false vacuum bubbles are singular.
If $M<0$, unlike the $M>0$ Schwarzschild geometry, 
the exterior geometry is globally static and there is no 
exterior horizon. One would not expect
to find inflating solutions, singular or otherwise. 
We will demonstrate explicitly that
one does not. 

When $\eta>M_p$, $\Delta\Omega > 4\pi$. 
If, in addition, $M<0$    
the causal roles of $R$ and $T_M$  in Eq. (\ref{eq:EXT}) are interchanged
beyond some finite value of $R$.
This value separates a static region from an expanding cosmological one.
This expanding region can support an inflating interior:
$r$ can  grow large without doing so at the expense of the 
monopole exterior.

While $\eta \gtrsim M_p$ is a necessary condition, it is 
not a sufficient condition.
We have already observed in Sec. II
that non-inflating solutions appear to
exist when $\eta \gtrsim M_p$  ---
they certainly do if $\eta$ is not too large
in the $\sigma\to 0$ limit. So 
$\eta \gtrsim M_p$ is clearly not a sufficient 
condition. We will demonstrate explicitly, 
however, that there is a critical value of $\eta$ 
above which all solutions inflate.

Whereas $F$ was positive in Minkowski space,  
its relevant general relativistic
extensions need not be: as demonstrated in 
Appendix B, $F_-$ 
assumes negative values along physical trajectories 
when the core is larger than a de Sitter horizon;
$F_+$ may be negative if the Barriola-Vilenkin geometry possesses
a horizon.

In this section, we will examine the solution of Eq. (\ref{eq:motion}).
The determination of the physical spacetime which corresponds to 
these solutions, and whether they are physically
realizable, will require the evaluation 
of the sign of $F_-$ and $F_+$. This  will be deferred to  
the next section.

It is convenient to introduce dimensionless variables.
We define

\bq 
        z= Hr\,,\quad 
        m  = {\sqrt{\lambda /3}\over 8\pi}{M\over M_p}\,,\quad
        \tilde \rho_\pm = {1\over \eta^2 H^2} \rho_\pm \,.
\label{eq:scale}
\nq
Let $\tilde \eta = \eta/M_p$ and
$\tilde\lambda = \lambda /3s^2$. 
We then have $F_\pm= H^{-2} {\cal F}_\pm$, where

\bq
{\cal F}_\pm = {1\over 2}\tilde\lambda^{1/2}\tilde\eta \left(m - 
{{\tilde\rho}_\pm\over 3} z^3 + z\right) \,,\qquad
           \tilde\rho_\pm = 3 (\tilde\eta^{-2} \pm  \tilde \lambda^{-1})\,.
\label{eq:calF}
\nq
The potential, $U$, is now parametrized by three 
dimensionless parameters: $\tilde\eta$ and $\tilde\lambda$
which characterize the physical theory, and 
the (reduced) gravitational mass, $m$. 

\bq
	U = - \l( {{\cal F}_- \over z^2} \r)^2 - z^2
	  = - \l( {{\cal F}_+ \over z^2} \r)^2 -{m \tilde\eta^2 \over z}
		-\tilde\eta^2\,.
\label{eq:calU}
\nq
We will consider a theory with a fixed value of $\tilde\lambda$
and examine the behavior as $\tilde\eta$ is dialled from $0$ to infinity.

Let us first examine the global properties of the  
function, $U$. As we will see, there are important respects in
which the functional form of $U$ differs from that
of its Minkowski space counterpart. 
To begin with, we note that $U\le 0$.
As $z\to 0$, 

\bq
           U\sim  - {m^2 \over z^4}\,,
\nq
independent of $G$. As $z\to \infty$
\footnote [2] {We note that $U \le - z^2$, with equality at points where 
${\cal F}_-=0$, if they exist.}, 

\bq
           U\sim - \left[\left(
                         {\tilde\rho_-\over 3}\right)^2 + 1 \right]\,z^2\,.
\nq
From its asymptotics,
it is clear that 
$U$ always possesses at least one positive maximum. Here we 
would like to determine the physical conditions 
under which it may possess additional positive critical points. 
Specifically, when does it possess a well? Recall that in 
Minkowski space 
this occurs when $M_c<M<0$, where $M_c$ is  
defined by Eq. (\ref{eq:Mc}).\footnote [3] 
{In analogy with the Minkowski space treatment, it is clear that 
a sufficient condition that it possesses two (or more) 
maxima is that ${\cal F}_-$ possesses two zeros. 
There is clearly then (at least) one minimum between them. 
This is not, however, a necessary condition.}

In general, at a critical point of $U$, $U'=0$ or

\bq
         {\tilde\lambda \tilde\eta^2\over 4} (z + m - {\tilde\rho_-\over 3}z^3)
                    (z+ 2m + {\tilde\rho_-\over 3}z^3) = z^6\,. 
\label{eq:U'}
\nq
We find that, for each $\tilde\eta$ 
there exists a critical negative mass, $m_c$, 
below which Eq. (\ref{eq:U'}) possesses a single 
positive solution, corresponding 
to a single maximum for $U$.
For negative $m> m_c$, the equation 
possesses three positive solutions, $z_-$, $z_0$ and $z_+$: 
two maxima, $z_-$ (left)
and $z_+$ (right), and a minimum $z_0$ lying between them.
As $m$ is lowered to $m_c$, the right hand maximum 
coalesces with the minimum and they annihilate. 
Let this value be $z_c$.  To determine $m_c$, 
we note that when $m=m_c$, in addition to $U'=0$, we also have  
$U''=0$ at this point. 
We can express these two conditions 
in the following form 

\bq
        m_c = {3 \over 2} {z_c \over 3-\tilde\rho_- z_c^2}
		\l( \tilde\rho_+^2 z_c^4 -3 \r)\,,
\label{eq:m2}
\nq
where
\bq
        \tilde\rho_+^2 (8 \tilde\rho_+^2 + \tilde\rho_-^2) z_c^8 
	- 18 \tilde\rho_- \tilde\rho_+^2 z_c^6 
	- 3 (\tilde\rho_+^2 - \tilde\rho_-^2) z_c^4 
	- 18 \tilde\rho_- z_c^2 - 9 =0\,.
\label{eq:z2}
\nq
There is a unique positive solution to Eq. (\ref{eq:z2}) for each
specification of $\tilde\eta$. 
Eq. (\ref{eq:m2}) then determines $m_c$.
This solution is given by the line labelled $m_c$ 
in the parameter space $(\tilde\eta,m)$
in Fig. \ref{fig=oscnvsm}. 
Clearly $m_c< m <0$ represents a necessary condition for a stable 
oscillating solution.

We have now determined how the number 
of critical points of the potential depends on $m$.

A necessary condition for classical motion is that $U\le -1$. 
For a given set of parameters, this condition will partition the 
domain of $U$ into allowed (or physical) and forbidden subdomains.
This partition, as well as
the qualitative nature of the motion in the disjoint physical domains,
is completely determined once we know 
where the critical points of $U$  
lie with respect to its turning points with $U=-1$.
Let us locate the mass values  at which critical points coincide with 
turning points.

The two equations, $U=-1$ and $U'=0$ 
can be reduced to a form analogous to 
(\ref{eq:m2}) and (\ref{eq:z2}):

\bq
        m= {z \over 3z^2 -2}
        \l[ 1-\l( 2 - {1 \over 3} \tilde\rho_- \r)
        z^2 \r]  \,,
\label{eq:m1}
\nq
where
\bq
        (z^2-1) (\tilde\rho_- z^2 -1)^2 
	+ {4 \over \tilde\lambda \tilde\eta^2}
   	z^2 (3z^2-2)^2 = 0\,.
\label{eq:z1}
\nq
The solutions are traced 
on the parameter space in Fig. \ref{fig=oscnvsm} where they are
labelled $m_0$, $m_+$
and $m_M$.

Let us describe these solutions in greater detail:

If $m_c<m<0$, we know that there are three critical points 
and the potential 
possesses a well. Under what conditions 
will a stable oscillating solution 
exists in this well? In general, we note that the left hand 
maximum is always the absolute maximum:

\bq
        U(z_-) > U(z_+)\,.
\nq
A stable oscillating solution will therefore exist if 
the motion is bounded by the right hand maximum, 

\bq
       U(z_+) > -1 \,,
\label{eq:up}
\nq
and the minimum lies within the physically accessible domain of $z$,

\bq     
       U(z_0) \le -1\,.
\label{eq:lower}
\nq
Eq. (\ref{eq:lower}) is saturated along $m_0$
on Fig. \ref{fig=oscnvsm}.
The mass spectrum of  stable oscillations is bounded from below by
$m_0$. There is  a single stationary solution of size $z_0$.

If $\tilde\eta <1$, Eq. (\ref{eq:up}) is always satisfied.
If $\tilde\eta\ge 1$, however, 
Eq. (\ref{eq:up}) is saturated along $m_+$.
Oscillating solutions of arbitrarily small negative mass exist when 
$\tilde\eta <1$ but do not when $\tilde\eta >1$.
If $\tilde \eta <1$, the mass spectrum is bounded from above by 
$m=0$; if $\tilde\eta\ge 1$, the spectrum is bounded from above by
$m_+$.

This behavior can be accounted for analytically. 
When $m \to -0$, the left maximum at $z_-$ and the minimum at $z_0$
degenerate to $z=0$ with $U(z_-)=0$ and $U(z_0) \to -\infty$.
In this limit, Eq. (\ref{eq:U'}) determines the position of the right 
hand maximum $z_+$,

\[
        \lim_{m\to -0} z_+^4 = {1 \over 4} 
                {\tilde\lambda \tilde\eta^2 \over
                1+{1 \over 36}\tilde\lambda \tilde\eta^2 \tilde\rho_-^2}\,.
\]
The condition, (\ref{eq:up}) then determines 
the limit, $\tilde\eta <1$.
This is independent of  $\tilde\lambda$. 
So for infinitesimally small and negative mass, oscillations
exist only when $\tilde\eta <1$.

Both $m_0$ and $m_+$ decrease monotonically with $\eta$. 
They terminate at the common point $P$ 
where they coincide with $m_c$.
This end point defines a critical value of $\tilde\eta$,
$\tilde\eta_c$,
above which there do not exist any stable  oscillating solutions. 
Wheareas in Minkowski space,
the boundary $m_c$  possesses no physical
significance, here it plays a role in the bifurcation in parameter space at
$\tilde\eta=\tilde\eta_c$.
We note that in the limit $\sigma\to 0$, 
$\tilde\eta_c\to \sqrt{3}$ consistent with the 
static limit of the de Sitter interior discussed in Sec. II. 
This, of course, is a highly singular limit.

There is no solution analogous to $m_+$ for $z_-$. 
This is because $U(z_-)> -1$ in this regime.

Both when $m<m_c$ and  $m>0$, there is a single maximum $z_M$
of $U$. The condition $U(z_M)= - 1$ identifies a mass, $m_M$.
See Fig. \ref{fig=oscnvsm}. 
If $m<0$ ($m>0$),  motion is monotonic 
when $m<m_M$ ($m>m_M$). $m=m_M$ therefore identifies
the boundary in parameter space separating regimes of monotonic 
and non-monotonic motion.
We note that along the positive branch, $m_M\to \infty$
as $\tilde\eta\to 0$.

We summarize as follows:

The boundaries $m_c$, $m_0$, $m_+$, $m_M$ and $m=0$
partition the parameter space 
into seven regions labelled (1) to (7)
in Fig. \ref{fig=oscnvsm}.

In the parametric region (1) bounded by 
$m=m_0$,  $m=m_+$ 
and $m=0$ 
the potential possesses oscillating, collapsing and expanding
domains. 
This  potential is plotted in Fig. \ref{fig=oscpot}.1. 

In regions (2)-(4), 
only collapsing and expanding domains exist.
In region (2), while the potential does possess a well, its
mimimum is inaccessible classically. See Fig. \ref{fig=oscpot}.2. 
There are no oscillatory solutions.
In region (3), the
potential again possesses a well. 
However, the right maximum no longer provides a 
barrier containing the motion in the well.
Motion in the well is unstable towards expansion.
See Fig. \ref{fig=oscpot}.3.
In region (4), the potential possesses  a single 
classically inaccessible maximum
(see Fig. \ref{fig=oscpot}.4) which separates 
collapsing from expanding domains.

In region (5), the single maximum of the potential 
is accessible.
The wall trajectory is monotonic (See Fig. \ref{fig=oscpot}.5).

For $m>0$, the potentials in region (6) and (7) are 
qualitatively identical to those
in region (4) and (5) respectively. See 
Fig. \ref{fig=oscpot}.6 and 
Fig. \ref{fig=oscpot}.7.
If $\tilde\eta <1$, the motion in these regimes is qualitatively 
identical to that of the false vacuum bubbles
discussed in Ref.\cite{BGG}.

In the next section, we will construct the 
spacetime which corresponds to each of these trajectories.

The above analysis simplifies 
considerably  if we assume $\tilde\rho_- = 0$ for all $\tilde\eta$ and 
$\tilde\lambda$.
As we will demonstrate in Appendix A, the overall picture is
qualitatively identical to the 
$s= {\rm constant}$ case.

\section{Embedding The Wall Trajectories in Spacetime}

In this section,
we trace the wall trajectory in spacetime. 

If $\tilde\eta <1$, 
the exterior metric (\ref{eq:EXT})
does not possess a  horizon. The singularity
at $R=0$ is a naked one. However, if we truncate the 
exterior spacetime at some finite value of $R$ and replace 
it by a non-singular patch of de Sitter space 
the singularity is removed.
The physical spacetime is free of singularities. 
The static coordinate system,
$(R,T_M)$, is globally valid in the exterior.  
$T_M$ is the asymptotic time and $\dot t_M$ must be 
positive along physical trajectories. 
Just as in the Minkowski space analysis, the sign of
$\dot t_M$ is the sign of $F_+$. So 
$F_+$ must be positive everywhere along trajectories.
In Appendix B. we show that $F_+$ is positive along the oscillatory
solution and negative along the others.
The only physical solutions are oscillatory.
It is simple to check that $z<1$ everywhere along 
these trajectories. The interior is completely 
covered by a static patch of de Sitter space.
It therefore does not inflate.
A remote observer will eventually see this stable 
oscillating motion of the wall. 
The remaining trajectories correspond to unphysical inside-out
solutions. This is completely analogous to the 
Minkowski space model discussed in Sec. III.

If $\tilde\eta >1$, the surface 
$R=r_H=2 G M/(1-\tilde\eta^2)$ is null. The coordinate system Eq. (\ref{eq:EXT})
breaks down at $R=r_H$. If $R<r_H$, the killing vector
$\partial_{T_M}$ is timelike, and the spacetime is static;
if $R>r_H$, $\partial_{T_M}$ is spacelike. The spacetime is
dynamical in this region. At large values of $R$, the spacetime metric
can be approximated by 

\bq
        ds^2 = 
       {-dR^2\over \tilde\eta^2 -1 } + 
       (\tilde\eta^2 -1 ) dT_M^2 + R^2 d\Omega^2\,.
\label{eq:EXT1}
\nq
As discussed in \cite{Cho},
each constant $T_M$ slice of the asymptotic geometry is a 
$2+1$ dimensional Friedman - Robertson - Walker universe
expanding linearly with time, $R$. 
The asymptotic geometry can be represented 
locally by a spherical cylinder $S^2\times R$.
This is a highly anisotropic cosmology.

We can introduce Kruskal-Szekeres coordinates, 
$(u,v)$ that cover the complete exterior spacetime. 
The maximal extension of this coordinate system  
is the spacetime shown in Fig. \ref{fig=oscks2}. This
spacetime is represented by a subset of a plane, to each point of 
which corresponds a two-sphere of radius $R$.
This plane divides naturally into four quadrants.
The coordinates are defined 
as follows in terms of the `static' coordinates
in each of these quadrants: 
in quadrants I and III (upper sign for quadrant  I),

\bqr
        u &=& \pm \l( 1-{R \over r_H} \r)^{1/2}\,\exp \l( {R \over 2r_H} \r)\,
                \cosh \l( {T_M \over t_1} \r)\,,\nonumber\\
        v &=& \pm \l( 1-{R \over r_H} \r)^{1/2}\,\exp \l( {R \over 2r_H} \r)\,
                \sinh \l( {T_M \over t_1} \r)\,;\nonumber
\nqr
in quadrants II and IV (upper sign for quadrant II),

\bqr
        u &=& \pm \l( {R \over r_H} -1 \r)^{1/2}\,\exp \l( {R \over 2r_H} \r)\,
                \sinh \l( {T_M \over t_1} \r)\,,\nonumber\\
        v &=& \pm \l( {R \over r_H} -1 \r)^{1/2}\,\exp \l( {R \over 2r_H} \r)\,
                \cosh \l( {T_M \over t_1} \r)\,,\nonumber
\nqr
where $t_1 = r_H^2/GM$.
We have

\[
        u^2-v^2 = \l( 1 - {R \over r_H} \r)\,\exp \l( {R \over r_H} \r)\,,
\]
\bq
        {v \over u}=\tanh \l( {T_M \over t_1} \r)\;({\rm in\;I\,,III})\,,
        \qquad  
        {u \over v}=\tanh \l( {R \over t_1} \r)\;({\rm in\;II\,,IV})\,.
\label{eq:tconst1}
\nq
The spacetime is defined by the region $u^2 -v^2 \le 1$
on the $u$-$v$ plane. $R$-constant lines are hyperbolas which 
degenerate into straight lines, $v=\pm u$ on the horizons,
$R=r_H$. There are two naked singularities 
given by the timelike hyperbolas, $u^2 -v^2 = 1$.
$T_M$-constant lines are straight lines passing through the origin.
 
One way to describe the global geometry of this spacetime is
to consider a foliation by constant $v$ hypersurfaces. 
The spacetime is symmetric with respect to time
reversal $v\to -v$. The hypersurface $v=0$ is 
momentarily static. The topology of this hypersurface is $S^3$
with singular poles. 
The equatorial radius is $r_H$.
As $v$ is increased, the equatorial radius increases
monotonically while the geodesic distance between the poles
increases more slowly. 
While the singularities of the maximally extended geometry 
are naked, they do not appear on the truncation of this
geometry which corresponds to the physical exterior region
unless the core originates at or collapses to $r=0$.

The physical exterior region surrounding a global monopole
is clearly very different from the 
global geometry we have just described.
Sakai {\it et al.} have described numerically the evolution of an
inflating  global 
monopole in a spacetime which is initially flat. 
The initial core radius in their model lies outside the monopole 
horizon.
Let us suppose that we place the monopole on the left of the $u-v$ plane.
The maximal extension of 
such a slice might be approximated by the flat spatial
hypersurface, $\Sigma$
on Fig. \ref{fig=oscks2}. It 
coincides with the hypersurface $v=0$ at the 
point $(-1,0)$, crosses the horizon $v=-u$ at some finite 
positive value of $v$ and tends 
asymptotically to $v=u$. 
The initial data 
on the slice will not generally be stationary.
The future of this slice lies entirely within the two quadrants
(II) and (III). The asymptotic region described by Eq. (\ref{eq:EXT1})
lies completely within region (II).
The remaining two quadrants
(I) and (IV) are inaccessible
for the boundary conditions we are considering. 
This is analogous to the description of 
stellar collapse, where half of the 
maximally extended Schwarzschild geometry 
gets discarded.

\begin{figure}
\begin{center}
\psfig{file=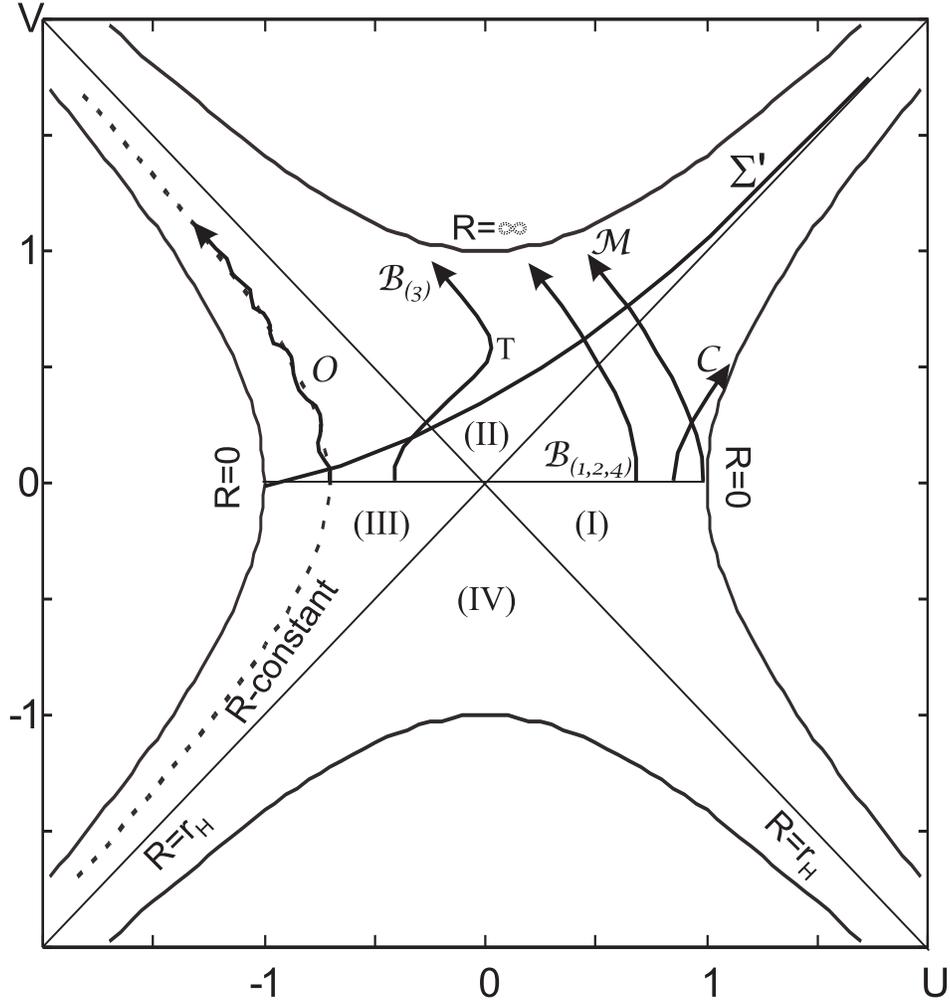}
\end{center}
\caption{
Trace of the wall trajectories on a Gibbons-Hawking diagram
for $\tilde\eta<\sqrt{\tilde\lambda}$.
The oscillating trajectory, ${\cal O}$, lies completely within
(III). The expanding trajectory, ${\cal B}_{(3)}$, 
lies completely within  (II) and (III). 
It changes its angular direction within (II).
The collapsing type ${\cal C}$ trajectory, as well as both  
expanding ${\cal B}_{(1),(2),(4)}$ and 
${\cal M}$ trajectories originate 
outside of the horizon.
${\cal B}$ and ${\cal M}$, however, cross the horizon and 
enter (II). 
${\cal B}_{(1),(2),(3),(4)} $ and ${\cal M}$
move asymptotically to the left.
}
\label{fig=oscgh}
\end{figure}

Now let us examine the interior de Sitter space. 
To trace the trajectory in de Sitter space it is 
convenient to describe it by Gibbons-Hawking
coordinates, $(U,V)$. 
These coordinates have been described elsewhere in detail. 
We refer the reader to \cite{BGG} for further details.
Briefly, with respect to these coordinates,
de Sitter space is represented by the region,
$|U^2-V^2| \le 1$ on the $U-V$ plane, each point of which 
represents a 2-sphere of radius $R$, related to $U$ and $V$ by
$U^2-V^2 = (1- H R)/ (1+ HR)$.
See Fig. \ref{fig=oscgh}. The de Sitter horizon of the north (south)
pole $(-1,0)$ ($(1,0)$) is represented by $V=U$ ($V=-U$).
The center of the monopole is located on the left at the north pole. 

The flat slice, $\Sigma$, in the 
maximal extension of the exterior 
geometry will be truncated at its 
intersection with the 
relevant wall trajectory. 
It can, however, be extended into the interior. 
Indeed, the standard presentation of de Sitter 
space is as a spatially flat Friedman-Robertson-Walker 
Universe. This slice is labelled $\Sigma'$
on  Fig. \ref{fig=oscgh}. $\Sigma'$ 
coincides with the hypersurface $V=0$ at the 
point $(-1,0)$, crosses the horizon $V=-U$ at some finite 
positive value of $V$ and tends 
asymptotically to $V=U$. 
The future of $\Sigma'$ lies entirely within the two quadrants
(II) and (III).
Initial data is defined on the union 
of the interior and exterior flat slices.

The role played by the two fugacities, $\beta_D$ and $\beta_M$,
was described in \cite{BGG} in the context of a 
Schwarzschild exterior. They demonstrated that
$\beta_D$ ($\beta_M$) was proportional to minus the angular velocity 
about the origin on a
Gibbons-Hawking (Krustal-Szekeres) spacetime diagram for the 
interior (exterior) respectively.  In Appendix B, we show that 
the sign of $\beta_D$ ($\beta_M$) coincides with the
sign of ${\cal F}_+$ (${\cal F}_-$).
These signs  therefore determine uniquely the 
routing of trajectories about the origins of 
Figs. \ref{fig=oscks2} and \ref{fig=oscgh}. 
The relevant interior and exterior
spacetimes can then be constructed with this information.

Let us first discuss the trajectories of the wall in the 
exterior.

All oscillating trajectories 
lie entirely within region (III) on the Krustral-Szekeres plane.
Their trajectories are labelled ${\cal O}$ in Fig. \ref{fig=oscks2}.

We describe bounces with the stationary initial condition, $\dot r=0$.

There are two qualitatively distinct expanding bounce trajectories.
The trajectory, labelled ${\cal B}_{(3)}$
in Fig. \ref{fig=oscks2} 
(which corresponds to  parameters lying within 
region $(3)$ in  Fig. \ref{fig=oscnvsm}), 
also originates in region (III) though this region 
need not be part of the physical spacetime. 
If it is, the exterior possesses a horizon.
The bounce may also then have a stationary point  
within the physical region.
It must, however, always cross the horizon and enter region (II).  
Its motion changes direction at some point outside the horizon 
which we can set to occur at $T_M=0$ without loss of generality.
(See Appendix B.) 
The  remaining expanding bounce trajectories,
${\cal B}_{(1),(2),(4)}$,  
(corresponding to parameters in regions (1), (2) and (4)
on Fig. \ref{fig=oscnvsm}), 
as well as all 
monotonically expanding trajectories, ${\cal M}$ 
(region (5)),
originate in the unphysical region (I).
They do not possess any physically accessible 
stationary points. However, they 
cross the horizon and enter the physical region (II). 
There is no horizon in the exterior.

All  collapsing trajectories, labelled ${\cal C}$
in Fig. \ref{fig=oscks2} 
lie  entirely within the unphysical region (I). 
They cannot be realized in an asymptotically
cosmological geometry.

As $\tilde\eta$ increases the horizon scale shrinks to zero, $r_H \to 0$,
where the exterior geometry is singular. 
The external metric (\ref{eq:EXT}) we have been 
exploiting, based on Barriola and Vilenkin's asymptotic exact solution,
is not expected to provide a valid approximation under these circumstances, 
In particular, in this limit, the description of the 
region $R<r_H$ breaks down. However, we also found that
oscillating solutions do not exist for large $\eta$;
the expanding solution of type ${\cal B}_{(3)}$ 
is expected to be formed in region (II) with 
an initial size much larger than $r_H$. 
The analysis of trajectories remains valid.

The expanding type ${\cal B}_{(3)}$ motion 
is unbounded, $r\to\infty$. However, 
outside the horizon, the exterior spacetime is itself expanding.
After an initial invasive period (to $T$ on Fig. \ref{fig=oscks2})
the expansion of the wall is not at the expense 
of the ambient exterior. All expanding trajectories 
move eventually to the left.  
Therefore, a remote observer in the exterior is safe not to be swallowed  
by the the expansion of the monopole.
This observer will, however, eventually see all 
trajectories.

In Figs. \ref{fig=oscgh} and \ref{fig=oscgh1}, 
we plot the corresponding 
embedding of possible wall trajectories in de Sitter space.

Oscillating trajectories, ${\cal O}$, 
both for $\tilde\eta \ge 1$ and  for $\tilde\eta <1$,
are contained completely within region III on the 
Gibbons-Hawking plane.
The wall does not cross the horizon and
the interior does not inflate.

Along the infinite trajectories of type ${\cal B}$ 
and ${\cal M}$, however, $r$ must cross the de Sitter horizon at some 
point and enter region II. The interior spacetime necessarily inflates.
The qualitative nature of these inflating solutions will
depend on the magnitude of $\tilde\eta$. There is a critical 
value of $\tilde\eta$, $\tilde\eta_D= \sqrt{\tilde\lambda}$,
below which all inflating trajectories move asymptotically to the 
left, and above which they all move to the right.
Compare Fig. \ref{fig=oscgh} with Fig. \ref{fig=oscgh1}.
Once any constant $T_D$ hypersurface is breached, it
cannot be recrossed. A foliation of de Sitter 
space by constant $V$ hypersurfaces
reveals that
the spherically symmetric inflating region 
in the former case ($\tilde\eta < \tilde\eta_D$)
is contained completely in the northern hemisphere of de Sitter
space; in the latter ($\tilde\eta > \tilde\eta_D$) it contains the equator.
The volume of the inflating region 
increases with increased $\tilde\eta$.

Technically, at the critical value, $\tilde\eta=\tilde\eta_D$,
$\tilde\rho_-=0$. If $\tilde\eta> \tilde\eta_D$, then $\tilde\rho_-<0$.
In appendix B., we demonstrate explicitly how
the sign of ${\cal F}_-$ and hence the routing of
trajectories is affected.
If $\tilde\eta<\tilde\eta_D$, ${\cal B}_{(3)}$ originates in (III)
changes direction 
outside the de Sitter horizon. ${\cal B}_{(1),(2),(4)}$ and ${\cal M}$
originate in (I). They do not change direction.  

If $\tilde\eta>\tilde\eta_D$,
${\cal B}_{(3)}$ still originates in (III). 
However, it no longer changes direction 
outside the de Sitter horizon. 
We note that $\tilde\eta_D> \tilde\eta_c$ so that
parameter regions (1) and  (2) are not present. 
${\cal B}_{(4)}$ now originates in (III). It does not change direction.
${\cal M}$ still originates in (I). However, 
it now changes direction in (II).

In general, 
the initial region of de Sitter 
space bounded by ${\cal B}_{(3)}$ 
may be smaller than the horizon. That bounded by
${\cal B}_{(4)}$ may also be when $\tilde\eta>\tilde\eta_D$.
${\cal B}_{(1),(2)}$ and ${\cal M}$, however,  always 
contain a horizon.


\section {conclusions}

We have presented a simple model of a global monopole. This 
model predicts that if $\tilde\eta \equiv \eta /M_p <1$,
there is always a stable 
static solution with a negative gravitational mass. There is 
no inflation. 
If $\tilde\eta >1$, but below some 
critical value, there exist classically 
stable analogues of these static monopoles. 
However, the exterior spacetime is no longer static.
For all $\tilde\eta>1$, there exist non-singular inflating monopoles with  
strictly negative gravitational mass.

These inflating solutions
differ from the false vacuum bubble solutions examined in \cite{BGG} in the
important respect that the spacetime is non-singular.

The monopole interior inflates when the core radius exceeds the 
de Sitter horizon radius. Once the wall radius, $r$, 
crosses the horizon it must continue to expand.
There are no inflating solutions with constant $r$.

In a false vacuum bubble with $M>0$, the inflating 
interior does not destroy the exterior
because it occurs behind the event horizon of the 
exterior Schwarzschild geometry.
In the present case with $M<0$, the picture is very different. There
are no horizons when $\tilde\eta<1$. But there is no inflation
so there is no difficulty reconciling physics
in the interior with that in the exterior.
However, when $\tilde\eta>1$ there are inflating 
monopoles but no wormhole and no event horizon.
To understand what is happening, we found it useful to 
exploit a Kruskal-Szekeres diagram, Fig. \ref{fig=oscks2} 
to represent the global spacetime structure of the exterior.
There is a cosmological horizon, $v=-u$, beyond 
which the spacetime to the future of any 
spacelike  asymptotically flat slice 
is dynamical. 
In this region the roles of $T$ and $R$ get interchanged.
The Schwarzschild parameter, $T$, is no longer the proper time of 
an inertial observer at $R=\infty$. 
$\partial_{T_M}$ is a spatial Killing vector in this region.
$R$ is time. As $R\to\infty$, 
the geometry  can be represented by a universe 
expanding linearly with time, $R$ along two directions.
Because of this expansion the unlimited expansion
of the core radius is not at the expense of the 
exterior spacetime. 
If $\eta$ is sufficiently large the monopole
horizon is small and  the core boundary 
will  always be located in the dynamical region. It inflates.

Considering the simplifying assumptions we have made to model
the system, the critical value we find,
 $\tilde\eta_c \approx 1.235$ ($\eta_c \approx 0.25m_p)$
agrees well with the value, 
$\eta_c \approx 0.33m_p$ determined numerically by Sakai {\it et al.}
\cite{NS}. In addition, we note that with a 
flatter $\phi^6$ potential, using the numerical
technique adapted in \cite{NS,Cho}
we obtain the numerical value,
$\eta_c \approx 0.265 m_p$
which is closer to our critical value.
The flatter the potential the better the thin wall approximation is
expected to be.

There are several open questions related to this work 
that merit further examination.

For each $\tilde\eta$ in the interval, $1 <\tilde\eta < \tilde\eta_c$
we saw that a stable oscillating solution 
co-exists with an expanding solution of the same mass in some  
strictly negative band of values of $m$. It would appear 
that all non-inflating monopoles are therefore metastable 
with $\tilde\eta >1$: there is the 
possibility of its tunneling into an inflating configuration.
In the semiclassical approximation tunneling 
will be described by an instanton which interpolates between the 
oscillating and inflating solutions \cite{FGG}.

A static monopole minimizes the energy within the 
topological class to which it belongs.
We would expect it to be stable against 
perturbations. It is not so clear what to expect in
an inflating monopole. This question can be addressed within the 
context of the present model by examining the 
stability of the exterior geometry with respect to 
perturbations. Formally this problem is almost identical to the 
analysis of perturbations about a Schwarzschild geometry.
However, the exterior is now cosmological, not static, and the boundary 
conditions involved are very different.

We have only considered global monopoles. It should be straightforward 
to examine gauge monopoles, as well as cosmic strings and 
domain walls in this approximation\cite{Z}.

\acknowledgments
\noindent{
This approach to global monopoles 
was suggested by Alexander Vilenkin and
we have benefitted greatly from conversations with
him. We also thank Daniel Sudarsky for 
helpful discussions. JG  gratefully acknowledge the hospitality of the 
Institute of Cosmology at Tufts University, support
from CONACyT Grant 211085-5-0118PE and a DGAPA 
sabbatical fellowship from UNAM.}

\appendix
\section{Analysis of the wall motion in a simplified limit}

In this section, we analyze the wall motion in the  analytically tractable
special case given by $\tilde\rho_-=0$.

We note that the reduced energy density difference, $\tilde\rho_-$, 
defined by (\ref{eq:calF}) is bounded from below
by the negative value, $-3\tilde\lambda^{-1}$.
In this regime, the gravitational potential is strong. It 
has no Minkowski space analogue. $\tilde\rho_-$ diverges to plus
infinity as $\tilde\eta\to 0$. This is the Minkowski limit.
$\tilde\rho_-$  vanishes when $\tilde \eta= \sqrt{\tilde\lambda}$.
Remarkably, the potential simplifies on this 
subset of parameter space deep in the non-perturbative regime.
When $\tilde\rho_-=0$,  

\bq
            {\cal F}_- = {1\over 2}\tilde\eta^2 (m + z)
\nq
is linear in $z$ with slope $\tilde\eta^2/2$.
The necessary condition for classical motion, 
$U\le -1$, now assumes the particularly simple form

\bq
          |{\cal F}_-| \ge {\cal G}_s(z) \,,
\label{eq:ineq}
\nq
where 

\[       
          {\cal G}_s(z) =  z^2 (1-z^2)^{1/2}
\]
if $z\le 1$ and zero otherwise.
We note that

\[
          {\cal G}_s'(z) = {z(2- 3z^2) \over(1-z^2)^{1/2}}\,,\qquad
          {\cal G}_s''(z) = { 2 - 9z^2 +6 z^4 \over (1-z^2)^{3/2}}\,.
\]
The function, ${\cal G}_s$, has a single maximum at
$z=\sqrt{2/3}$, an inflection point at 
$z_i=\sqrt{(9-\sqrt{33})/12}$, and vanishes at $z=0$ and $z=1$. 
The inequality (\ref{eq:ineq}) is very simply illustrated graphically.
There are three possibilities.
Two intersections of the graphs $|{\cal F}_-|$ and ${\cal G}_s$ 
indicate the existence of two bounce motions 
described in the text, one of which collapses the other expands.
Four intersections (three with ${\cal F}_-$ and one with $-{\cal F}_-$)
indicate, in addition, the existence of 
an oscillating solution  lying in the domain between the two bounces.
Zero intersection indicates a monotonic solution. We illustrate all
three possibilities in Fig. \ref{fig=oscfgs}.

With $\tilde\eta$ fixed, let us dial $m$ beginning with $m=0$ 
and decreasing through negative values.

Suppose that $m$ is small and negative. 
Only two of the possibilities described above are possible:
either there are two or there are four interections of $|{\cal F}_-|$ 
and ${\cal G}_s$. Now when $m=0$, $-{\cal F}_-$ can intersect 
${\cal G}_s$ only at $z=0$.
${\cal F}_-$ also intersects ${\cal G}_s$ at $z=0$. It will re-intersect 
${\cal G}_s$ only if its slope lies below 
some critical value. We find

\[
         z^4 - z^2 + \left({\tilde\eta^2\over 2}\right)^2 \le 0\,.
\]
There exist two real solutions if and only if the discriminant of the 
quadratic in $z^2$ is positive, or 
$\tilde\eta\le 1$. This reproduces the criterion for the
existence of a zero mass oscillating solution obtained  
earlier in our general discussion.

Let $\tilde\eta>1$. It is clear that if, 
in addition, $\tilde\eta$ lies above some 
critical value $\tilde\eta_c$, ${\cal F}_-$ will never intersect ${\cal G}_s$ 
more than once.
This critical value occurs when
the slope of ${\cal F}_-$ exceeds the maximum slope of ${\cal G}_s$.
Thus $\tilde\eta_c^2 = 2 {\cal G}_s'(z=z_i)$, where  

\[
         {\cal G}_s'(z=z_i) = {1\over 4} (-1 + \sqrt{33}) \sqrt{ {9 -\sqrt{33}
                                 \over 3 + \sqrt{33}}}\,,
\]
which gives
 $\tilde\eta_c \approx 1.203$.

For all $\tilde\eta$ within the range $1\le \tilde\eta\le \tilde\eta_c$,
there will exist some band $[m_0,m_+]$ within which 
${\cal F}_-$ will intersect ${\cal G}_s$ three times. 
These values are determined by the real solutions of

\bq 
         {\cal F}_- = {\cal G}_s\,,\quad {\cal F}_-' = {\cal G}_s'\,,
\label{eq:FGs}
\nq
satisfying respectively ${\cal G}_s''<0$, and ${\cal G}_s''>0$.
These equations (\ref{eq:FGs}) reduce to 
Eq. (\ref{eq:z1}) and (\ref{eq:m1}) on setting $\tilde\rho_-=0$. 
The solution is represented by 
the lines $m_0$, $m_+$
on the $\tilde\eta - m$ plane in Fig. \ref{fig=oscfgs}. The 
interpretation is identical to that for the generic case.

A simple bound can be placed on $m_M$  by inspection. 
Let the position of intersection of $-{\cal F}_-$ and  ${\cal G}_s$ 
be $z_M$ when $m=m_M$. We have
$m_M = -z_M - 2 {\cal G}_s(z_{_M})\tilde\eta^{-2}$. 
An upper bound is obtained by approximating the position 
of intersection by $z=1$, $m_M< -1$. 
A lower bound is obtained by replacing ${\cal G}_s(z_M)$ by the maximum of
${\cal G}_s$ which is $2\sqrt{3}$. Thus

\bq
         m_M \ge  - 1 - {4\over 3\sqrt{3}} \tilde\eta^{-2}\,.
\nq 
The one qualitative difference with the generic case 
is that 
$m_M \to -\infty$ as $\tilde\eta \to 0$
along its negative branch.
This divergence can be justified as follows:
if $\tilde\rho_- =0$, the limit
$\tilde\eta \to 0$ implies that $s\to \infty$.
A large value of $s$ corresponds to a 
large energy density on the surface compared to 
that in the interior --- a situation
which is never realized in a monopole.

Finally, we can determine the 
functional form,  $m_c = m_c(\tilde\eta)$ exactly. 
Setting $\tilde\rho_-=0$, Eq. (\ref{eq:z2}) implies 

\[
        z = \l( {1+\sqrt{33} \over 48}\,{\tilde\lambda \tilde\eta^2 \over 4}
                \r)^{1/4}
          = \l( {1+\sqrt{33} \over 192} \r)^{1/4}
                \tilde\eta\,.
\]
Eq. (\ref{eq:m2}) then determines 

\[
        m(\tilde\eta) = {-15+\sqrt{33} \over 48} \l({1+\sqrt{33} \over 12}
                \r)^{1/4} \tilde\eta\,,
\]
which is linear in  $\tilde\eta$ with negative slope, and
independent of Newton's constant, $G$.

\section{Routing of the wall trajectories on Kruskal-Szekeres and
Gibbons-Hawking diagrams}

Let us rewrite Eq. (\ref{eq:calU})
as

\[
        \dot{z}^2 = { 1 \over z^4}
                \left( {\cal F}_\pm^2 - {\cal G}_{M,D} \right) \,,
\]
where 
\[
	{\cal G}_M = z^3 \l[ (1-\tilde\eta^2) z - m\tilde\eta^2 \r]\,,
\quad 	{\cal G}_D =z^4 (1-z^2)\,,
\]
and ${\cal F}_{\pm}$ are given by
Eq. (\ref{eq:calF}).
Classical motion of the wall is allowed
only in domains where 
${\cal F}_+^2 > {\cal G}_{M}$ or, equivalently,
${\cal F}_-^2 > {\cal G}_{D}$.
Points where ${\cal F}_\pm^2 = {\cal G}_{M,D}$, if they exist,
mark the turning points of the motion.

In Sec. V we claimed that the sign of ${\cal F}_+$ and 
${\cal F}_-$ 
determine the routing of the wall 
trajectory on the Kruskal-Szekeres and Gibbons-Hawking diagrams
respectively. To show this (in the former case), we note that 
Eq. (\ref{eq:Theta}) - (\ref{eq:U+}) relate ${\cal F}_+$ to
the fugacity, $\beta_M$: ${\cal F}_+= z^2 \beta_M$.
We exploit the definition of $\beta_M$,
and Eq. (\ref{eq:tconst1}) to give 

\[
        \beta_M = A_M \dot t_M =
                {8G^2M^2 \over (1-8\pi G\eta^2)^2}\, {1 \over r}\,
                \exp \l( -{1-8\pi G\eta^2 \over 2GM}\,r \r)
                (\dot{u}v-u\dot{v})\,.
\] 
This is independent of the quadrant in question. 
The change of the polar angle, $\theta_M=\tan^{-1} (v/u)$,
on the Kruskal-Szekeres plane is

\[
        \dot\theta_M = {u\dot{v}-\dot{u} v \over u^2} \cos^2\theta_M\,.
\]
Therefore $\beta_M \sim -\dot\theta$. 
The sign of ${\cal F}_+$ determines the routing of the trajectory about 
the origin, $(0,0)$. 
Positive ${\cal F}_+$ corresponding to clockwise motion.

In de Sitter space, we get

\[
        \beta_D = -A_D \dot t_D =
                -{1 \over H} (1+Hr)^2(U\dot{V}-\dot{U}V)
                \sim -\dot\theta_D\,,
\]
where $\theta_D=\tan^{-1} (V/U)$ and 
${\cal F}_-= z^2 \beta_D$. As before,  
the sign of ${\cal F}_-$ determines the routing of the 
trajectory on the Gibbons-Hawking diagram. Positive
${\cal F}_-$, as for ${\cal F}_+$  implies closewise
motion.

\begin{figure}
\begin{center}
\psfig{file=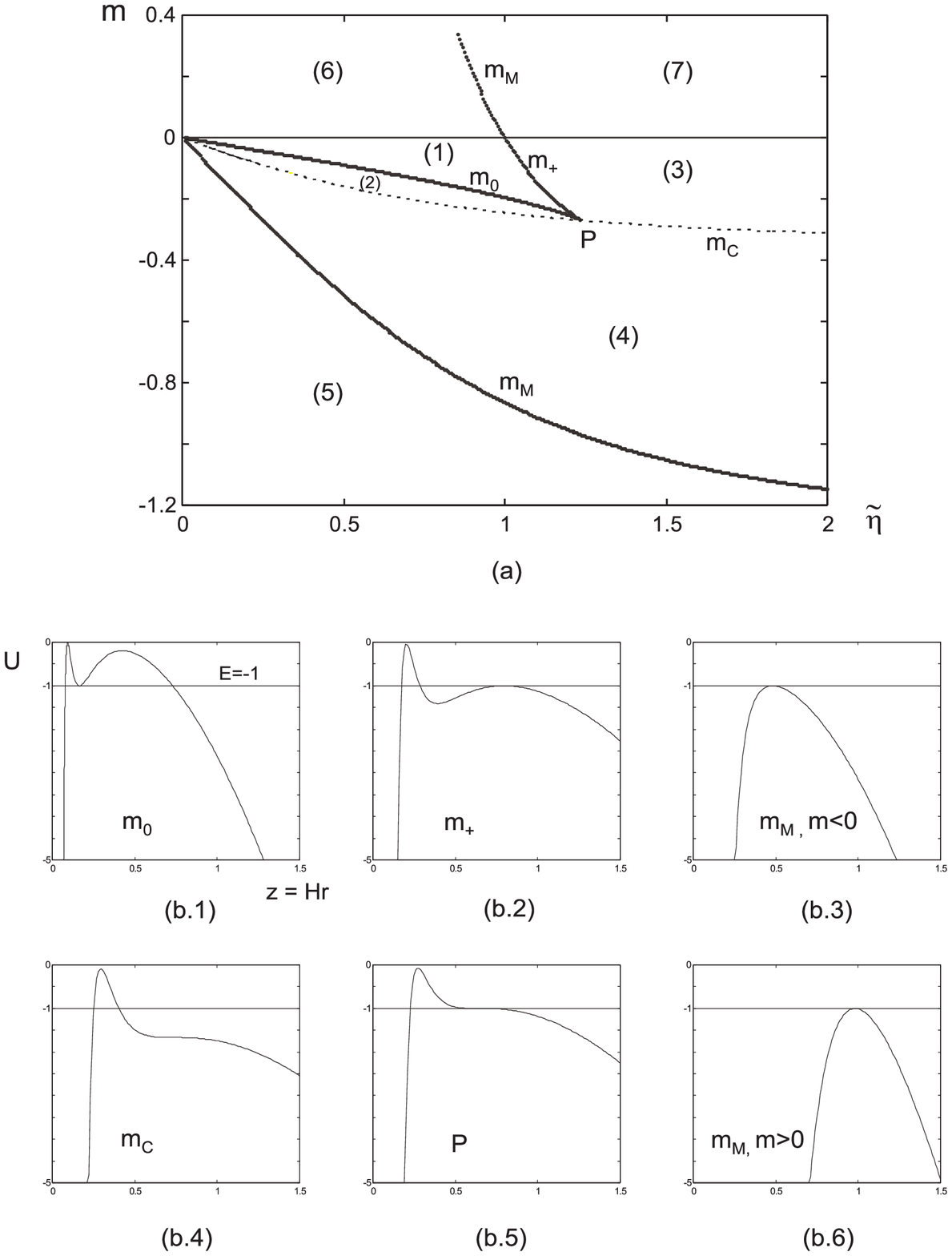}
\end{center}
\caption{
(a) The parametric space of the model, $m$ vs. $\tilde\eta$ with
$\tilde\lambda (\lambda=0.1\,,s=0.1)= 3.33$. In the region $m_c<m<0$ 
the potential, $U$, possesses two maxima 
$z_-$, $z_+$,  and a minimum, $z_0$ between them. 
Along $m_0$, $U=-1$ at $z_0$; along $m_+$, $U=-1$ at $z_+$. 
In the regions $m<m_c$ and $m>0$, $U$ has a single maximum, $z_M$.
$U=-1$ at $z=z_M$ along $m_M$. 
$m_0$ and $m_+$ terminate at the point, 
$P$ which lies on $m_c$.
The boundaries  
$m_c$, $m_0$, $m_+$, $m_M$ and $m=0$
partition the parameter space into seven regions with
qualitatively different potentials.
The potentials on these boundaries are plotted 
in Fig. \ref{fig=oscnvsm}(b).
(b.1) $m_0$ ($\tilde\eta=0.5\,,\;m=-0.0896$),
(b.2) $m_+$ ($\tilde\eta=1.15\,,\;m=-2.006$),
(b.3) $m_M$, $m<0$ ($\tilde\eta=0.5\,,\;m=-0.5136$),
and (b.4) $m_c$ ($\tilde\eta=1.57\,,\;m=-0.2926$).
(b.5) At P ($\tilde\eta=\tilde\eta_c=1.235\,,\;m=-0.2677$),
(b.6) $m_M$, $m>0$ ($\tilde\eta=0.5\,,\;m=2.9212$) .
}
\label{fig=oscnvsm}
\end{figure}

\begin{figure}
\begin{center}
\psfig{file=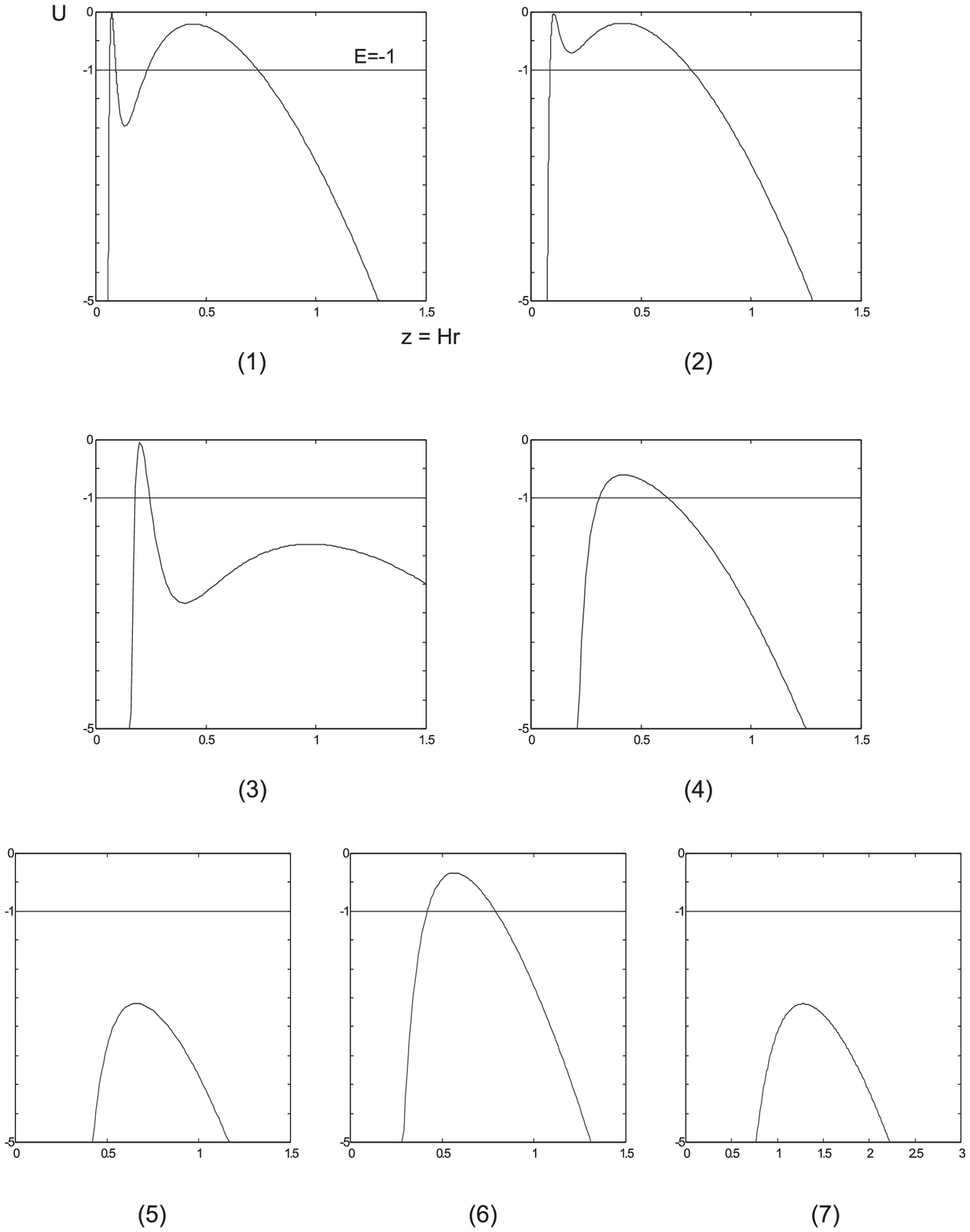}
\end{center}
\caption{
Plots of the potential in each of the seven 
regions of Fig. \ref{fig=oscnvsm}. 
(1)  ($\tilde\eta=0.5\,,\;m=-0.07$),
(2)  ($\tilde\eta=0.5\,,\;m=-0.1$),
(3)  ($\tilde\eta=1.5\,,\;m=-0.2$),
(4)  ($\tilde\eta=0.5\,,\;m=-0.4$),
(5)  ($\tilde\eta=0.5\,,\;m=-1$),
(6)  ($\tilde\eta=0.5\,,\;m=0.2$),
(7)  ($\tilde\eta=1.5\,,\;m=0.2$). 
The potential in region (1) possesses oscillating,
collapsing and expanding domains.
In region (2), the potential has a well but oscillations are 
not allowed classically.
In region (3), the right maximum of the potential does not contain the 
motion.
In region (4) and (6), the potential has a single maximum separating
collapsing and expanding domains.
In region (5) and (7), the single maximum is accessible. 
The motion is monotonic.
}
\label{fig=oscpot}
\end{figure}

\begin{figure}
\begin{center}
\psfig{file=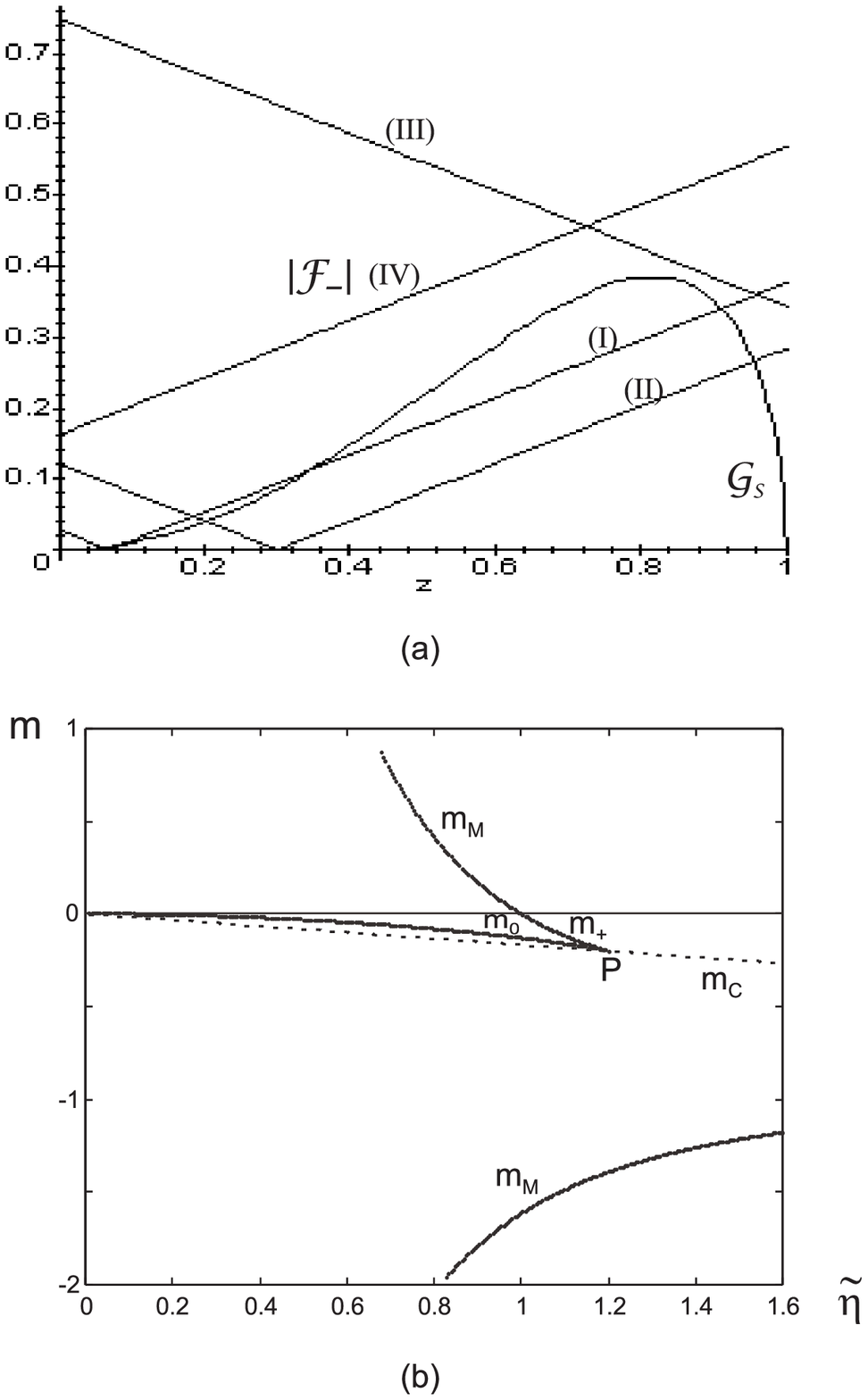}
\end{center}
\caption{
(a) Plots of $|{\cal F}_-|$ vs. $z$ and ${\cal G}_s$
vs. $z$ with fixed $\tilde\eta=0.9$.
With a small negative mass (I: $m=-0.07$), there are four intersections 
corresponding to bouncing points in the potential.
As $m$ decreases, oscillation disapears (II: $m=-0.3$) and then the wall
trajectory becomes monotonic (III: $m=-1.85$) as in the case of
positive mass (IV: $m=0.4$). 
(b) Parameter space with $\tilde\rho_-=0$ (Compare
Fig. \ref{fig=oscnvsm}(a)) 
The only qualitative divergence  from 
the generic behavior is  that $m_M$ diverges to $-\infty$
as $\tilde\eta\to 0$.
}
\label{fig=oscfgs}
\end{figure}

\begin{figure}
\begin{center}
\psfig{file=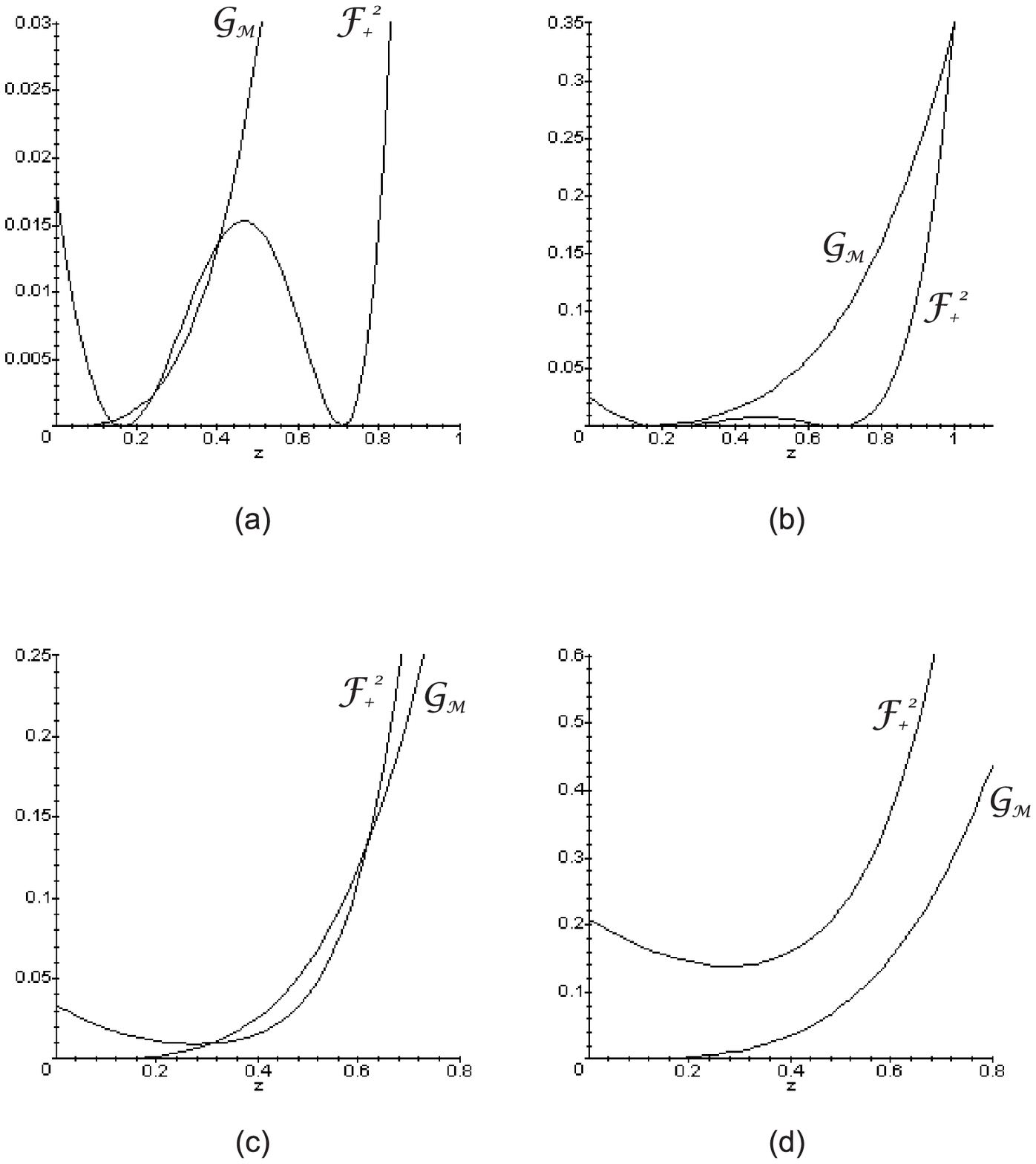}
\end{center}
\caption{
${\cal F}_+^2$ vs. $z$ and ${\cal G}_M$ vs. $z$ for $\tilde\eta < 1$
in the four regions of parameter space: (1), (2), (4), and (5)
(see Fig. \ref{fig=oscnvsm}).
(a) region (1) ($\tilde\eta=0.9\,,\;m=-0.16$),
(b) (2) ($\tilde\eta=0.9\,,\;m=-0.2$),
(c) (4) ($\tilde\eta=0.5\,,\;m=-0.4$),
and (d) (5) ($\tilde\eta=0.5\,,\;m=-1$).
${\cal F}_+$ is positive in the central physical domain of 
${\cal F}_+^2$ in (a) and (b) while it is negative in the other two.
Intersections of ${\cal F}_+^2$ and ${\cal G}_M$ indicate 
the turning points of the potential.
In (a), the last intersection on the top right is not shown.
}
\label{fig=oscfgm1}
\end{figure}

If $\tilde\eta < 1$, there are four
distinct regimes in parameter space we need to analyse,
labelled (1), (2), (4) and (5) in Fig. \ref{fig=oscnvsm}.
We plot both ${\cal F}_+^2$ and ${\cal G}_M$ vs. $z$
in Fig. \ref{fig=oscfgm1} for each of these cases in turn. 
${\cal G}_M$ is positive everywhere because there
is no horizon in the exterior. As we argued in Sec. V.
the only physically acceptable values of ${\cal F}_+$ are positive.

\noindent In (1), ${\cal G}_M$ intersects ${\cal F}_+^2$ at 4 points.
There are 3 disjoint domains where classical motion is allowed.
From the left to the right,
they correspond to collapsing, oscillating and expanding motion.
The sign of ${\cal F}_+$ is positive on the oscillating 
domain and negative in the others. 

\noindent In (2), 
${\cal G}_M > {\cal F}_+^2$ in the central region, indicating 
that the minimum of the potential
well is inaccessible  --- there are 
no classically allowed oscillations. 
However, ${\cal F}_+$ is negative on the 
two accessible domains.

\noindent 
In (4) ${\cal G}_M$ intersects ${\cal F}_+^2$; in (5) it does not
(motion is monotonic).
${\cal F}_+$ is negative everywhere in both cases.

\begin{figure}
\begin{center}
\psfig{file=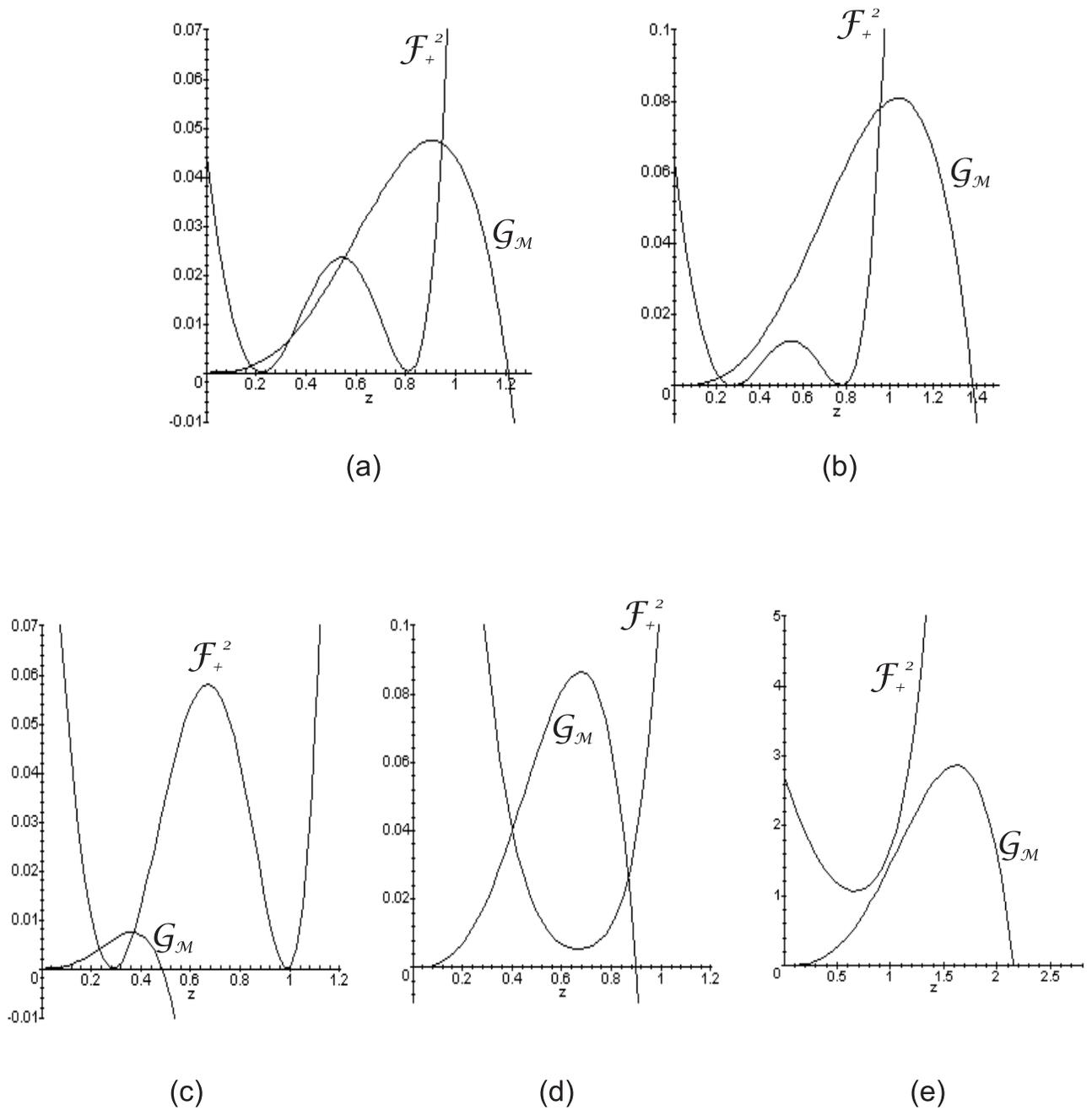}
\end{center}
\caption{
${\cal F}_+^2$ and ${\cal G}_M$ vs. $z$ for $\tilde\eta > 1$ in the 
five regions, (1) to (5) of  Fig. \ref{fig=oscnvsm}:
(a) region (1) ($\tilde\eta=1.1\,,\;m=-0.21$),
(b) (2) ($\tilde\eta=1.1\,,\;m=-0.25$),
(c) (3) ($\tilde\eta=1.5\,,\;m=-0.27$),
(d) (4) ($\tilde\eta=1.5\,,\;m=-0.5$)
and (e) (5) ($\tilde\eta=1.5\,,\;m=-1.2$).
${\cal F}_+$ changes sign on the expanding domain in (c).
${\cal G}_M$ vanishes at $z=0$ and at the exterior 
horizon. 
}
\label{fig=oscfgm2}
\end{figure}

If $\tilde\eta >1$, 
${\cal G}_M=0$ when $R=r_H$.
Whereas Fig. \ref{fig=oscfgm2} is superficially identical 
to Fig. \ref{fig=oscfgm1}
for  the regimes (1), (2), (4) and (5),
the physical interpretation is very different
because of the existence of the horizon.

\begin{figure}
\begin{center}
\psfig{file=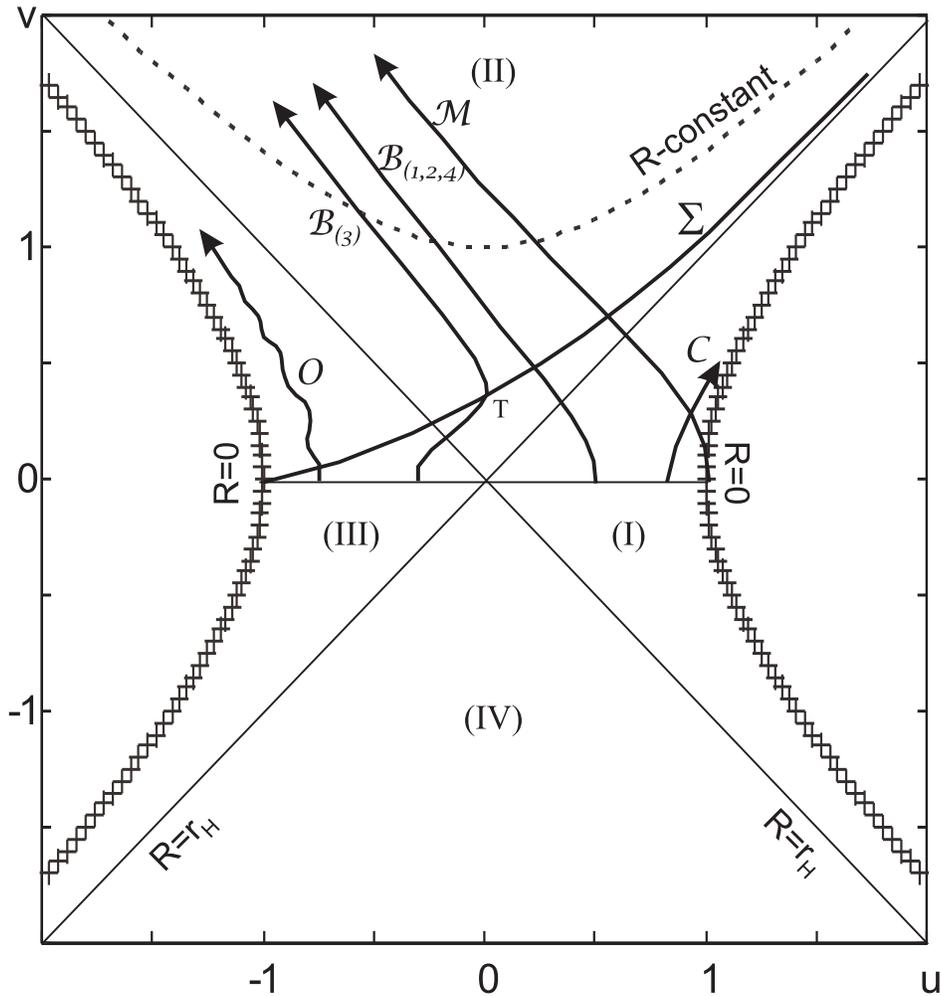}
\end{center}
\caption{
Trace of the wall trajectories on a Kruskal-Szekeres diagram for
$\tilde\eta > 1$. The physical 
exterior lies in region (II) and (III).
An identical physically inaccessable copy is 
provided by the regions (I) and (IV).
The oscillating trajectory, ${\cal O}$ (region $(1)$) lies within the
static region (III).
The expanding bounce trajectory, ${\cal B}_{(3)}$ 
also originates there but crosses the null surface
$u=-v$ to enter region (II), 
changing its angular direction at $T$. The trajectory
lies completely inside the physical region.
The collapsing trajectory ${\cal C}$ lies within the 
unphysical region (I). 
The remaining bounces ${\cal B}_{(1),(2),(4)}$
and the monotonically expanding ${\cal M}$ 
trajectories originate in the 
unphysical region but 
cross the horizon to enter (II). 
}
\label{fig=oscks2}
\end{figure}

\noindent In (1), 
${\cal F}_+$ is positive on the oscillating 
domain. This locates the oscillating trajectory in (III) and not in (I).
${\cal F}_+$ is negative in the remaining domains. The collapsing
trajectory necessarily lies in (I). The expanding solution 
has its stationary point in (I), crosses the horizon and enters (II).
See Fig. \ref{fig=oscks2}.

\noindent In (2) and (4), the collapsing and expanding trajectories are 
located as in (1) above. In (5), the monotonic trajectory 
originates at $r=0$ in region (I), 
crosses the horizon and enters (II) exactly like the expanding 
trajectories in (2) and (4).

All intersections of ${\cal F}_+^2$ with ${\cal G}_M$
lie  within the horizon, $r<r_H$. Stationary points necessarily lie in the 
static region. 

\noindent In the remaining regime (3), there is an additional feature 
we have not encountered so far. 
In the expanding domain, ${\cal F}_+$ changes sign from positive to
negative beyond the horizon.
This means that the trajectory 
(labelled ${\cal B}_{(3)}$)
must change its angular course at some point 
on the Kruskal-Szekeres plane. Without loss of generality we can always set 
this value to $T_M=0$. This trajectory evolves closewise 
prior to this point and counter-clockwise thereafter.
See Fig. \ref{fig=oscks2}.

Now let us turn inwards.
In contrast to ${\cal F}_+$, the sign of the
leading term, $-\tilde\rho_-z^3$ appearing in 
${\cal F}_-$, does  depend on the magnitude of $\eta$.

\begin{figure}
\begin{center}
\psfig{file=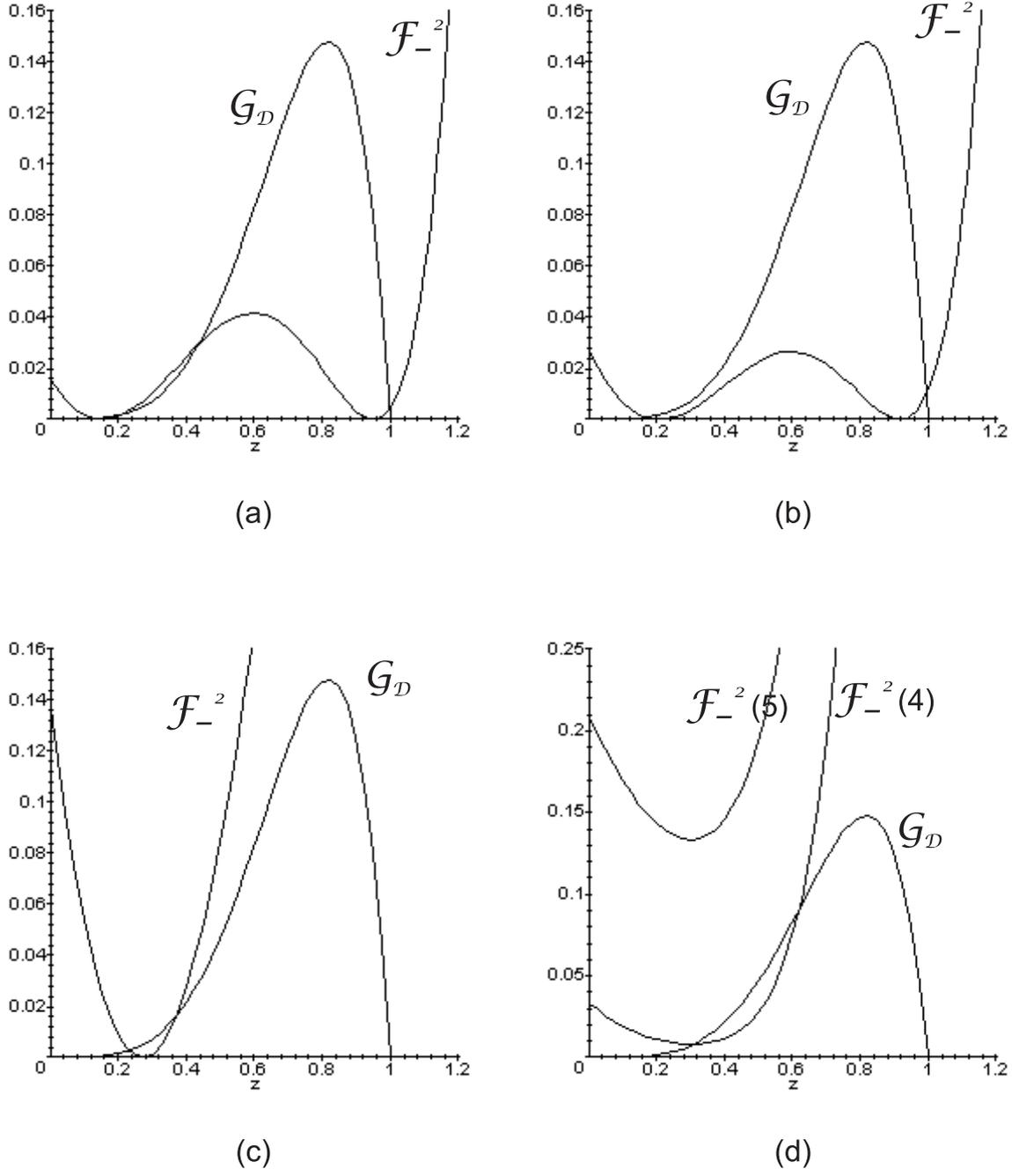}
\end{center}
\caption{
${\cal F}_-^2$ and ${\cal G}_D$ vs. $z$
for $\tilde\eta < \sqrt{\tilde\lambda}$.
(a) region (1) ($\tilde\eta=0.9\,,\;m=-0.15$),
(b) (2) ($\tilde\eta=0.9\,,\;m=-0.2$),
(c) (3) ($\tilde\eta=1.5\,,\;m=-0.27$),
(d-4) (4) ($\tilde\eta=0.5\,,\;m=-0.4$)
and (d-5) (5) ($\tilde\eta=0.5\,,\;m=-1$).
${\cal F}_-$ is positive on the central physical domain 
and negative in the others.
In (c), the right part of ${\cal F}_-^2$ is not shown.
}
\label{fig=oscfgd1}
\end{figure}

If $\tilde\rho_- > 0$ ($\tilde\eta < \sqrt{\tilde\lambda}$), 
the functional form  of
${\cal F}_-$ is qualitatively identical
to that of ${\cal F}_+$ for $\tilde\eta >1$
and the location of
trajectories on the Gibbons-Hawking plane  the same as that for
the Kruskal-Szekeres plane (Fig. \ref{fig=oscfgd1}).

\begin{figure}
\begin{center}
\psfig{file=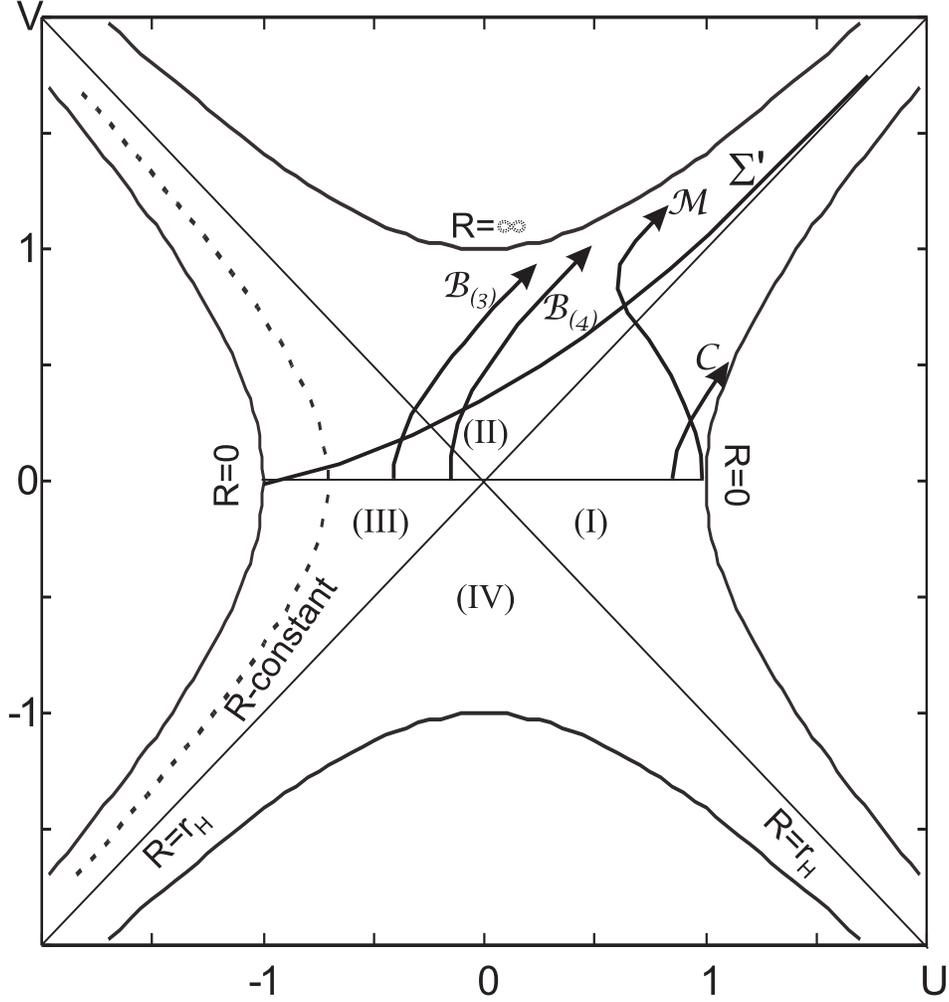}
\end{center}
\caption{
Trace of the wall trajectories on a Gibbons-Hawking diagram
for $\tilde\eta>\sqrt{\tilde\lambda}$.
The expanding trajectories, ${\cal B}_{(3)}$ and ${\cal B}_{(4)}$, 
lie completely within  (II) and (III). 
The collapsing type ${\cal C}$ trajectory as well as  
expanding  ${\cal M}$ trajectory originate 
outside of the horizon.
${\cal M}$, however, crosses the horizon and 
then changes its angular direction in (II). 
${\cal B}_{(3),(4)} $ and ${\cal M}$
move asymptotically to the right.
}
\label{fig=oscgh1}
\end{figure}

\begin{figure}
\begin{center}
\psfig{file=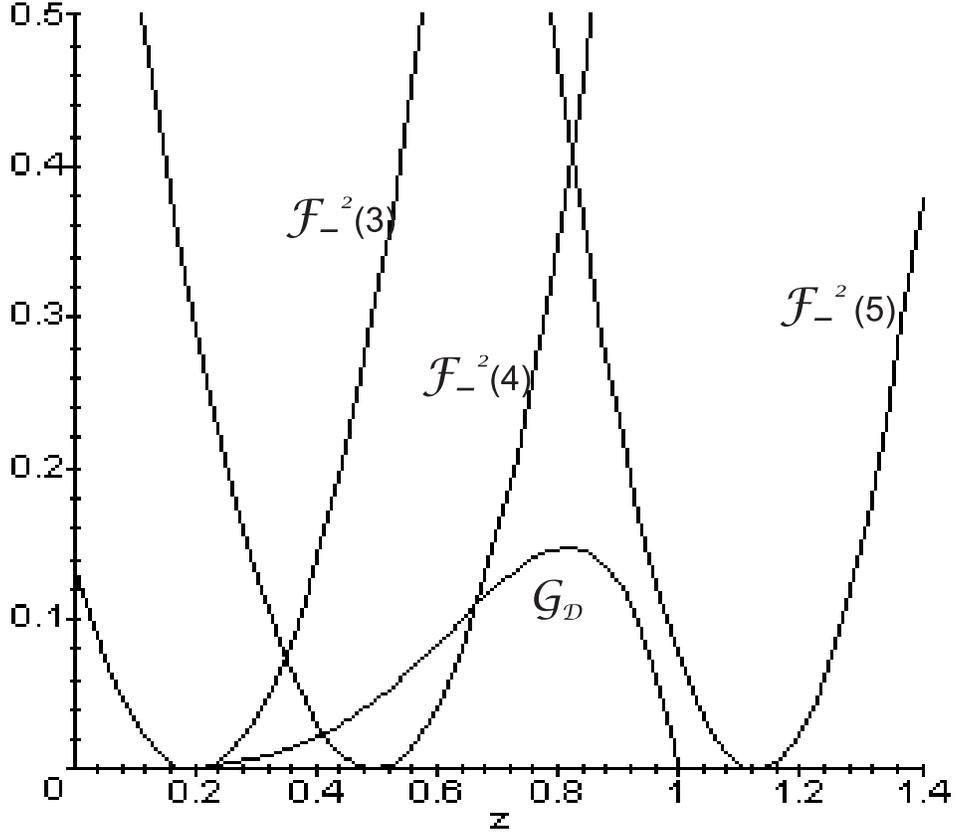}
\end{center}
\caption{
${\cal F}_-^2$ and ${\cal G}_D$ vs. $z$ 
for $\tilde\eta > \sqrt{\tilde\lambda}$
in region (3) ($\tilde\eta=2\,,\;m=-0.2$), 
(4) ($\tilde\eta=2\,,\;m=-0.5$)  and
(5) ($\tilde\eta=2\,,\;m=-1.2$).
${\cal F}_-$ is positive in the domain of increasing ${\cal F}_-^2$.
In region (5), ${\cal F}_-$ changes sign in the expanding domain.
}
\label{fig=oscfgd2}
\end{figure}

If $\tilde\rho_- < 0$ ($\tilde\eta > \sqrt{\tilde\lambda}$), 
however, ${\cal F}_-^2$ intersects ${\cal G}_D$ twice in regions (3) and (4)
(Fig. \ref{fig=oscfgd2}).
Unlike the previous case, ${\cal F}_- > 0$ everywhere along 
the expanding ${\cal B}_{(3)}$ trajectories. The polar angle is
monotonic. The stationary point of ${\cal B}_{(4)}$ has moved from 
(I) into (III). See Fig. \ref{fig=oscgh1}. 
${\cal F}_-$, on the other hand, 
now changes sign in the domain, $z>1$ along
${\cal M}_{(5)}$. The corresponding trajectories traced
on Fig. \ref{fig=oscgh1}.

\newpage

\newpage

\end{document}